\documentclass[useAMS,usenatbib]{mn2e}

\DeclareSymbolFont{cmletters}{OML}{cmm}{m}{it}
\DeclareMathSymbol{v}{\mathalpha}{cmletters}{"76}

\usepackage{amsmath}
\usepackage{amssymb}
\usepackage{epsfig}
\usepackage{graphicx}
\usepackage{ifthen}
\usepackage{latexsym}
\usepackage{rotating}
\usepackage{subfigure}
\usepackage{times,epsf}
\usepackage{txfonts}
\usepackage{varioref}
\usepackage{verbatim}
\usepackage{url}
\usepackage{color}
\usepackage[dvipsnames]{xcolor}
\usepackage[T1]{fontenc}
\usepackage{float}

\newcommand{\be}{\begin{equation}}
\newcommand{\ee}{\end{equation}}
\newcommand{\bea}{\begin{eqnarray}}
\newcommand{\eea}{\end{eqnarray}}

\title[Local Stability of Strongly Magnetized Black Hole Tori]{Local Stability of Strongly Magnetized Black Hole Tori}
\author[M. Wielgus et al.]
       {Maciek Wielgus$^{1, 3}$\thanks{E-mail: maciek.wielgus@gmail.com},
	P. Chris Fragile$^{2}$,
        Ziming Wang$^{2}$,
        and Julia Wilson$^{2}$\\
        $^1$ Nicolaus Copernicus Astronomical Center, ul. Bartycka 18, PL-00-716 Warszawa, Poland\\
        $^2$ Department of Physics and Astronomy, College of Charleston, Charleston, SC 29424, USA\\
        $^3$ Institute of Micromechanics and Photonics, ul. \'Sw. A. Boboli 8, PL-02-525 Warszawa, Poland}

\begin{document}

\maketitle

\label{firstpage}

\begin{abstract}
We investigate the local stability of strongly magnetized relativistic tori orbiting Kerr black holes, for the case of a purely toroidal magnetic field topology. Our approach encompasses both solving the full dispersion relation numerically and solving a simplified analytic treatment. Both approaches indicate that such tori are subject to an unstable non-axisymmetric magnetorotational mode, regardless of such background properties as the gas-to-magnetic pressure ratio and specific angular momentum distribution. We demonstrate that our modal analysis matches well with what is seen in global, three-dimensional, general relativistic magnetohydrodynamic simulations.  
\end{abstract}


\section{Introduction}
\label{introduction}

Recent decades have seen an evolution in our comprehension of the role of magnetic fields within astrophysical accretion discs.  Early discussions on the topic generally considered discs to be solely fluid flows, in which the magnetic fields were assumed to be dynamically weak and insignificant to stability \citep[e.g.][]{Abramowicz78a}. Nevertheless, even some of these early studies recognized that magnetohydrodynamic (MHD) turbulence could potentially alter, or even be the source of, the disc viscosity that allows matter to flow into the black hole \citep[e.g.][]{Shakura73}. This hypothesis was finally confirmed with the application of the magnetorotational instability (MRI) to accretion flows \citep{Balbus91}; today, the MRI is recognized as the primary source of turbulence that drives accretion within most discs.

Despite this recognition, most studies have only considered the case of weak magnetic fields; measured in terms of the ratio of gas pressure, $P_g$, to magnetic pressure $P_m$, they have assumed $\beta = P_g/P_m \gg 1$.  However, numerical simulations have shown that $\beta$ often approaches 1 close to the central object and away from the midplane of the disc \citep{DeVilliers03, McKinney04, McKinney06}. Furthermore, there is indirect evidence that magnetic fields must be strong in at least some regions of accretion flows.  For instance, dynamically strong, poloidal magnetic field are thought to be necessary for powering the relativistic jets that are often associated with black hole accretion disc systems \citep{Blandford77, DeVilliers05, McKinney06}.  

One possible way to introduce strong magnetic fields into an accretion disc is through a thermal runaway in a disc that initially has a weak magnetic field \citep{Pariev03, Machida06,Fragile09}.  Runaway cooling in an initially hot, thick, weakly magnetized disc will cause the thermal pressure in the disc to drop, which will, in turn, cause the disc to collapse vertically.  Being weak, the magnetic field is trapped by the gas, while magnetic flux is conserved.  To achieve this conservation, the magnetic field strength goes up as the cross-sectional area of the disc decreases.  Again, this is happening while the gas pressure within the disc is declining, so the ratio of gas pressure to magnetic pressure, or $\beta$, can drop considerably in such a scenario.  

In this work we will focus on strong toroidal magnetic fields.  The role of toroidal magnetic fields in discs have been considered in a few previous studies.  Among the important discoveries were that the existence of a toroidal field may significantly reduce the growth rate of many instabilities compared to cases where only weak vertical fields are considered \citep{Balbus91, Blaes94} and may even stabilize the disc against certain instabilities \citep{Knobloch92}. Furthermore, \citet{Kim00} concluded that when the magnetic field strength is superthermal ($\beta < 1$), the inclusion of toroidal fields tends to suppress the growth of the MRI, and that for quasi-toroidal field configurations, no axisymmetric MRI takes place in the limit of the sound speed $c_s \to 0$. In fact, when strong magnetic fields are considered, the presence of a~toroidal component plays a crucial role not only in determining the growth rates of the unstable modes but also in determining which modes are subject to instabilities \citep{Pessah05}. 

We re-explore this issue through the use of linear perturbation numerical and analytic calculations as well as global, three-dimensional numerical simulations of strongly magnetized tori. We start from the exact equilibrium analytic solution of \citet{Komissarov06}, that generalizes the construction of a thick disk \citep[a.k.a. ``Polish donut'';][]{Abramowicz78a} by admitting the presence of a~toroidal magnetic field. We find that the torus is susceptible to a~single unstable, local, non-axisymmetric mode, which we identify as a non-axisymmetric MRI mode \citep{Balbus92}. 

Throughout this paper, we use the natural units $G=c=M=1$ and $(-+++)$ as the spacetime signature. We also use units where factors of $4\pi$ are absorbed into the definitions of the magnetic field (Lorentz-Heaviside units). The paper is organized as follows: In Section \ref{s.model} we briefly discuss the construction of the Komissarov solution, including its extension to non-constant specific angular momentum. In Section \ref{s.perturb} we consider a local, linear perturbation analysis of the MHD equations for the Komissarov torus. Then, in Section \ref{s.numeric} we present results of global, general relativistic MHD numerical simulations of the same. We contrast stable two-dimensional evolution with unstable three-dimensional evolution.  Finally, in Section \ref{s.discuss}, we offer our closing thoughts on magnetized tori stability.

\section{Analytic model of Komissarov}
\label{s.model}

In the framework of general relativity, the evolution of an ideal, magnetized fluid can be described by the covariant equations of ideal MHD \citep{Dixon78}
\begin{equation}
\nabla_\alpha T^{\alpha \beta} = 0 ~,
\label{eq:ConserveT}
\end{equation}
\begin{equation}
\nabla_\alpha \left(^\ast F^{\alpha\beta}\right) = 0 ~, 
\label{eq:Maxwell}
\end{equation}
\begin{equation}
\nabla_\alpha \left(\rho u^{\alpha}\right) = 0 ~,
\label{eq:ConserveM}
\end{equation}
where $T^{\alpha\beta}$ is the total energy-momentum tensor of the fluid and electromagnetic field \citep{Anile89}
\begin{equation}
T^{\alpha\beta} = (w + b^2) u^{\alpha} u^\beta + \left( P_g +  \frac{1}{2} b^2 \right) g^{\alpha\beta} - b^\alpha b^\beta ~,
\end{equation}
$^\ast F^{\alpha\beta} = u^\alpha b^\beta - b^\alpha u^\beta$ is the dual of the Faraday tensor, $\rho$ is the gas rest mass density, $u^\alpha$ is the fluid four-velocity, $w$ is the specific enthalpy, ${P_g}$ is the gas pressure, and $b^2 = b^{\mu} b_{\mu}$ is twice the magnetic pressure $P_m$.  

\citet{Komissarov06} provides a class of steady-state solutions to equations (\ref{eq:ConserveT})-(\ref{eq:ConserveM}), generalizing the hydrodynamical torus construction of \citet{Abramowicz78a}. These solutions yield a~model of a~gas torus threaded by a purely toroidal magnetic field. While the construction is valid for any stationary, axisymmetric spacetime, we choose to work in the Kerr spacetime hereafter. We further assume purely rotational fluid motion and purely toroidal magnetic field topology, i.e.,
\begin{equation}
u^r = u^ {\theta} = b^r = b^{\theta} = 0 ~.
\end{equation}
Under the additional assumption that either the specific angular momentum $\ell = -u_{\phi}/u_{t}$ is constant or the angular velocity $\Omega = u^{\phi}/u^{t}$ can be expressed as a function of $\ell$
\begin{equation}
{\Omega} = - \frac{ g_{t\phi}+ \ell g_{t t}} {g_{\phi\phi}+\ell g_{t \phi}} \equiv \Omega(\ell) \ \text{,}
\label{eq:OmegaEll}
\end{equation}
the following equation can be derived from equations (\ref{eq:ConserveT})-(\ref{eq:ConserveM}) \citep{Komissarov06}:
\begin{equation}
\ln{\mid{{u_t}}\mid} - \ln{\mid{u_{t_{in}}}\mid} - \int^{\ell}_{\ell_{in}}{\frac{\Omega d\ell}{1 - \Omega \ell} + \int^{P_g}_{0}{\frac{dP_g}{w}} + \int^{\tilde{P_m}}_{0}{\frac{d\tilde{P_m}}{\tilde{w}}}}= 0 \ \text{.}
\end{equation}
Here, $\tilde{P_m} = \mathcal{L} {P_m}$ and $\tilde{w} = \mathcal{L}w$, with the relativistic term $\mathcal{L} = g_{t\phi}g_{t \phi} - g_{tt}g_{\phi \phi}$. This equation demands that the gas and magnetic pressures vanish at the torus surface, where $\ell = \ell_{in}$ and $u_t = u_{t_{in}}$. Assuming the state equations
\begin{equation}
P_g = K_gw^\Gamma ~,
\end{equation}
\begin{equation}
\tilde{P_m} = K_m\tilde{w}^\Theta ~,
\end{equation} 
we finally reach the formula
\begin{equation}
\ln{\mid{{u_t}}\mid} - \ln{\mid{u_{t_{in}}}\mid} + \frac{\Gamma}{\Gamma-1} \frac{P_g}{w} + \frac{\Theta}{\Theta-1} \frac{P_m}{w} =  \int^{\ell}_{\ell_{in}}\frac{\Omega d\ell}{1 - \Omega \ell} ~,
\label{eq:KomSol}
\end{equation}
which can also be expressed as
\begin{equation}
\Delta W + \frac{\Gamma}{\Gamma-1} \frac{P_g}{w} + \frac{\Theta}{\Theta-1} \frac{P_m}{w} =  0
\label{eq:KomSol2}
\end{equation}
for the introduced total potential $W$. Satisfying this equation is all that is required of our magnetized torus solution.

The \textit{center of the torus} is located at $(r_{c},\pi/2)$, where $r_{c}$ is the larger of two radii for which the specific angular momentum $\ell(r_c, \pi/2)$ is equal to the local Keplerian value, see Fig. \ref{fig:AngMomEquat}. At this point we parametrize the magnetic field strength in terms of the pressures ratio
\begin{equation}
\beta_c = \frac{P_g(r_{c},\pi/2)}{P_m(r_{c},\pi/2)} ~.
\end{equation}
Note that the total (gas + magnetic) pressure can not be arbitrarily large in the considered model, i.e., strong magnetic field in the context rather means ``strong magnetic pressure \textit{compared to} the gas pressure (small value of $\beta$)''.

Thus, the minimum set of parameters to specify the Komissarov solution uniquely is
\begin{enumerate}
\item black hole mass $M$ and spin $a$ to determine the underlying spacetime geometry,
\item $\ell(r, \theta)$ to fix the geometry of the eigenpotential surfaces,
\item $\Delta W = W_{in} - W_c$ to determine the size of the torus,
\item $\beta_c$ to parametrize the magnetic field.
\end{enumerate}
Throughout this paper we keep $\Gamma = \Theta = 4/3$ fixed. 

The \citet{Komissarov06} solution is important since it constitutes a rare case of an analytic solution in relativistic magnetohydrodynamics.  Despite its simplicity, it has been used in direct astrophysical applications, such as investigating the spectral properties of Sgr A* \citep{Yan14}. It also constitutes an important benchmark for numerical MHD solvers, since in 2D it has been demonstrated to be stable under axisymmetric perturbations \citep{Komissarov06}. In this paper, however, we find that this stability does not extend to the case when non-axisymmetric perturbations are considered, even with strong (i.e., superthermal) magnetic fields.

\subsection{Non-constant Specific Angular Momentum}

While in the simulations \citet{Komissarov06} only considered the simplest case of the constant specific angular momentum, $\ell = \ell_0$, when the right hand side of equation (\ref{eq:KomSol}) vanishes, the construction allows for non-constant specific angular momentum as well. The necessary condition is that the angular velocity be expressible as a~function of the specific angular momentum $\Omega = \Omega(\ell)$. Note that this condition is always fulfilled for pure rotation with a~barotropic equation of state, which is the case for the Komissarov torus solution\footnote{This follows from the relativistic von Zeipel theorem; both $\ell$ and $\Omega$ are constant on the same class of surfaces, the von Zeipel cylinders.}. We extend the original set of solutions slightly, by considering the angular velocity to be a~power law function of $\ell$
\begin{equation}
\Omega(\ell) = c_0 \ell^{n_0} ~,
\end{equation}
for which the right hand side of the equation (\ref{eq:KomSol}) can be evaluated to give
\begin{equation}
\int^{\ell}_{\ell_{in}}\frac{\Omega d\ell}{1 - \Omega \ell} = \frac{1}{n_0+1}\ln \left( \frac{ c_0 \ell_{in}^{n_0+1} - 1}{c_0 \ell^{n_0+1} - 1 }\right) ~.
\label{eq:IntegralPowerOmega}
\end{equation}
In order to determine the distribution $\ell(r, \theta)$, we first calculate $c_0$ and $n_0$ using the two assumed radii $r_{\rm cusp}$ and $r_{c}$, at which the angular momentum $\ell(r, \pi/2)$ in the equatorial plane is equal to its Keplerian value. Then we evaluate $\Omega(r_{\rm cusp})$ and $\Omega(r_c)$ with the general formula \ref{eq:OmegaEll} and find parameters $c_0$ and $n_0$. Then the $\ell(r,\theta)$ distribution can be found by solving numerically
\begin{equation}
c_0 \ell^{n_0} + \frac{ g_{t\phi}+ \ell g_{t t}} {g_{\phi\phi}+\ell g_{t \phi}} = 0
\end{equation}
for $\ell$ on the grid of $(r, \theta)$. Subsequently, the remaining quantities describing the magnetized torus solution can be found with equations (\ref{eq:KomSol}) and (\ref{eq:IntegralPowerOmega}).

\begin{figure}
   \centering
   \includegraphics[width=0.48\textwidth]{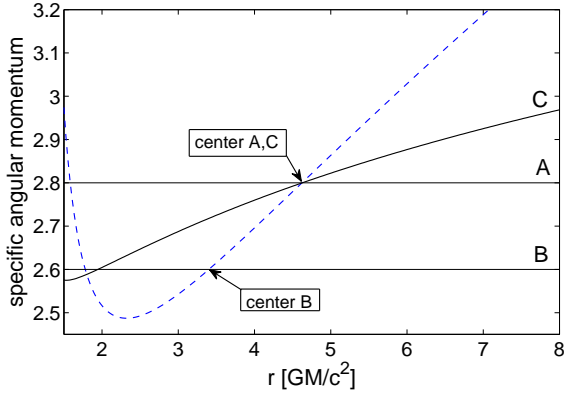}
   \caption{Radial distribution of specific angular momentum in the equatorial plane for the three cases considered. The Keplerian distribution for spin $a = 0.9$ is shown with the {\em dashed} line, for comparison. Locations of the tori centers are indicated.}
   \label{fig:AngMomEquat}
\end{figure}

\begin{figure}
   \centering
    \includegraphics[trim = 0 0mm 0mm -6mm, width=0.435\textwidth]{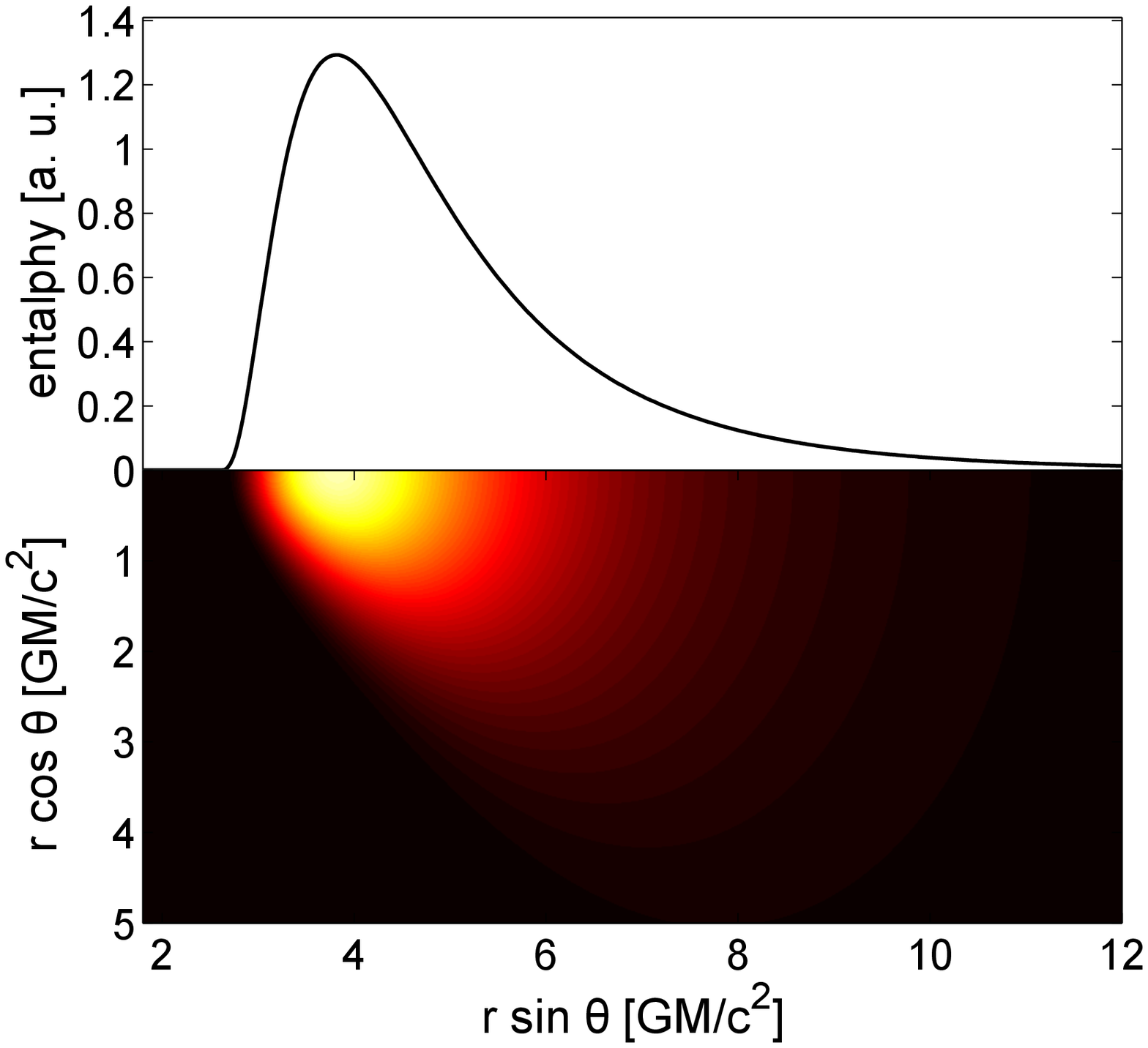}
   \includegraphics[trim = 0 0mm 0mm 0mm, width=0.435\textwidth]{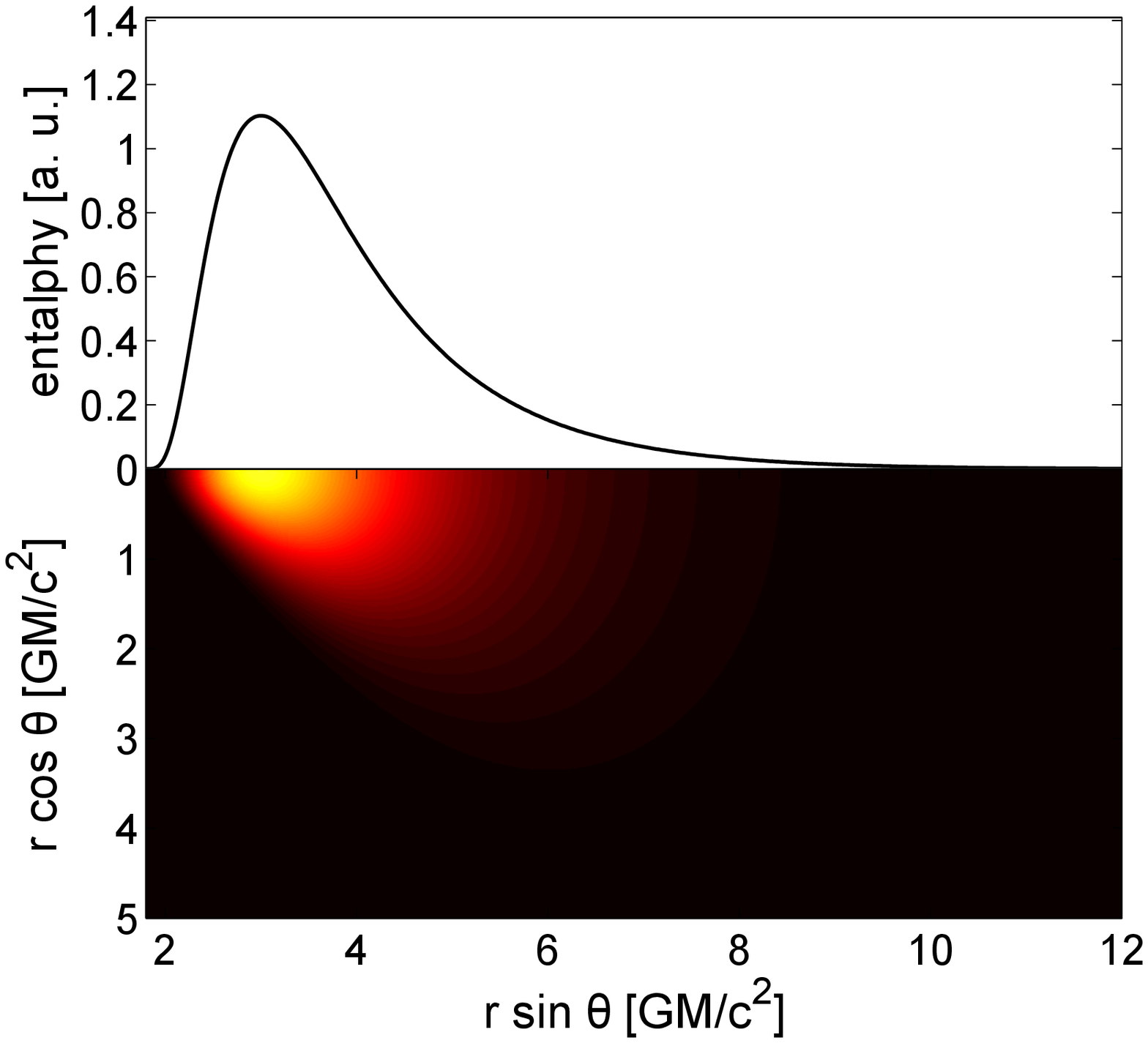}
   \includegraphics[trim = 0 0mm 0mm 00mm, width=0.435\textwidth]{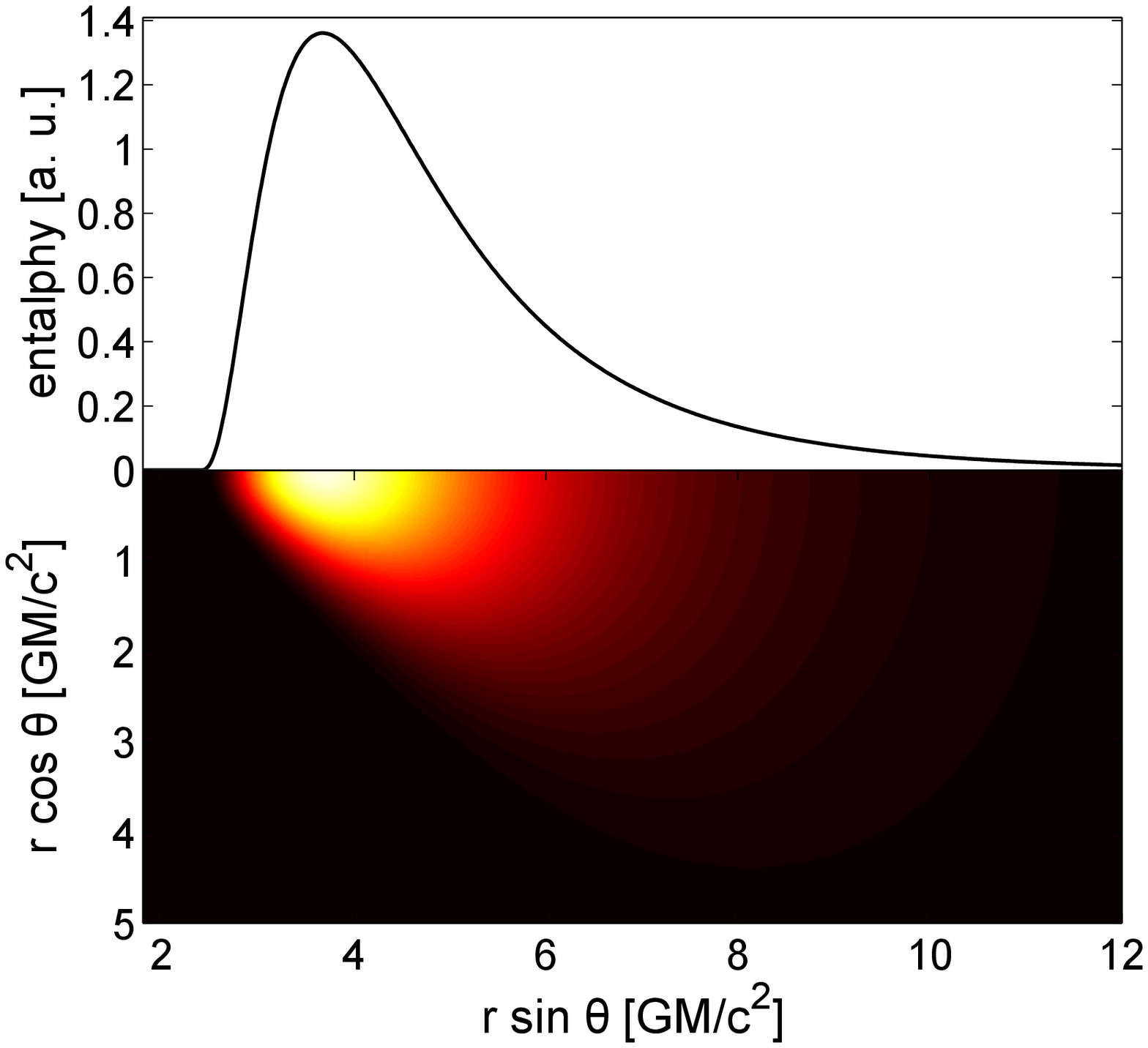}
      \caption{Enthalpy distribution for the considered tori: line plot of the enthalpy distribution in the equatorial plane ({\em top half} of each panel) and pseudocolor image of enthalpy in the $(r, \theta)$ plane ({\em bottom half} of each panel). {\em Top} panel: case A; {\em Middle} panel: case B; and {\em Bottom} panel: case C (non-constant specific angular momentum).}
   \label{fig:KomissarovPL}
\end{figure}

We consider three particular tori in this paper, denoted by letters A, B, and C. Cases A and B are the same as the ones presented in \citet{Komissarov06}. Case C is a non-constant specific angular momentum case (see Fig. \ref{fig:AngMomEquat}). 
All three are calculated for a Kerr black hole with dimensionless spin parameter $a = 0.9$. The remaining parameters for each case are listed in Table \ref{tab:params}.
{\setlength{\tabcolsep}{5.1pt}
\begin{table}    
   \begin{tabular}{|c|l|l|l|l|l|l|l|l|}
   \hline
    Case & $l_0$ & $r_{\rm cusp}$ & $r_c$ & $W_{\rm cusp}$ & $W_{c}$ & $W_{in}$ & $\Delta W$ & $\beta_c$ \\
   \hline
     A & 2.8 & 1.58 & 4.62 &  0.702 & -0.103 & -0.030 & 0.073 & 0.1 \\  
   \hline
     B & 2.6 & 1.78 & 3.40 & -0.053 & -0.136 & -0.053 & 0.083 & 1.0 \\ 
   \hline
     C & $\cdots$ & 1.80 & 4.62 & 0.017 & -0.103 & -0.048 & 0.055 & 0.1 \\ 
   \hline
   \end{tabular}
\caption{Parameters of considered solutions}
\label{tab:params}
\end{table}}
Figure \ref{fig:KomissarovPL} shows the enthalpy distributions $w(r, \theta)$ for each of the cases. 
Note that in the presence of a strong magnetic field, the entalphy maximum \textit{does not} coincide exactly with the location of the torus center $r_c$.

\section{Local perturbative analysis}
\label{s.perturb}

We perform a local linear stability analysis of the Komissarov torus configuration. Our Eulerian perturbative approach is limited to the equatorial plane, with all vertical derivatives $\partial_\theta$ assumed to vanish. Otherwise, the approach is quite general, since we perturb the fully relativistic equations (\ref{eq:ConserveT})-(\ref{eq:ConserveM}) and do not employ further simplifications such as the Boussinesq approximation. It is important to do so, since we consider highly relativistic tori solutions; one can not hope to find quantitatively correct results using Newtonian analysis in only few mass radii distance from the singularity. One caveat, though, is that, while we do consider non-axisymmetric perturbations, we neglect the shearing background effect \citep{Goldreich65}, which requires more in-depth analysis. Comments on that issue are given in Section \ref{sect:shearingBckg}.

\subsection{Full system of equations}
\label{s.full}

For a given physical quantity $n$, the perturbation is assumed to be of the form 
\be
\underline{\delta  n} = \delta n \exp \left[ - i (\omega t + k_r r + k_\theta \theta + k_\phi \phi) \right] ~,
\ee
where $\delta n$ is a constant complex number (the amplitude). Note that $k_\theta$ and $k_\phi$ have the dimension of rad$^{-1}$. The quantities we perturb are the fluid four-velocity, $u^\alpha$, the magnetic field, $b^\alpha$, and the enthalpy $w$. Thus, there are 9 perturbed variables in total. Additionally, the amplitudes must satisfy the following constraints, derived from the four-velocity normalization
\be
u_\alpha u^\alpha = -1 \implies \delta u^t = \ell \delta u^\phi ~,
\label{eq:4velocityNorm}
\ee
and the orthogonality of $b^\alpha$ and $u^\alpha$
\be 
u^\alpha b_\alpha = 0 \implies b_\alpha \delta u^\alpha + u^\alpha \delta b_\alpha = 0 ~.
\label{eq:uborthogonal}
\ee
We utilize equation (\ref{eq:4velocityNorm}) to reduce the number of variables to 8. The final perturbed linear system to be solved consists of 3 Euler equations (time, azimuthal, vertical), 3 induction equations (time, azimuthal, vertical), the orthogonality condition as represented by equation (\ref{eq:uborthogonal}), and the continuity equation. Note that in our convention the unstable modes correspond to \emph{positive imaginary part} of $\omega$. The full system is given in Appendix \ref{ap.system}. Deriving the dispersion relation demands rather exhaustive use of algebra and is prone to errors, hence we evaluate it numerically. Since the perturbed system is linear in $\omega$, it can be written as
\be 
\mathbf{S}(S_0, \omega, k_i) = \mathbf{A}(S_0) \omega + \mathbf{B}(S_0, k_i) ~,
\ee
where $S_0$ denotes the underlying steady state solution and $\mathbf{A}$, $\mathbf{B}$, and $\mathbf{S}$ are $8 \times 8$ matrices. Eigenfrequencies can be found as generalized eigenvalues of the pair of matrices $\mathbf{A}$, $\mathbf{B}$, i.e., as solutions to the problem
\be 
\mathbf{S} \delta \mathbf{v}^{(j)} = \left[ \mathbf{A} \omega^{(j)} + \mathbf{B} \right] \delta \mathbf{v}^{(j)} = 0
\ee
for some eigenvector $\delta \mathbf{v}^{(j)}$, or simply, solutions to
\be 
D_{(S_0,k_i)}(\omega) \equiv \det \left[  \mathbf{S}(S_0, \omega, k_i) \right] = 0 ~,
\label{eq:DispersionRelation}
\ee
for a given $S_0$ and wave vector $k_i$. Equation (\ref{eq:DispersionRelation}) is simply a compact form of writing the dispersion relation $D(\omega)$, which is a 7th order polynomial function of $\omega$ (since one of the 8 equations does not depend on $\omega$). Stability is determined by the signs of the imaginary parts of the roots of $D(\omega)$, i.e., $\omega^{(j)}$ for $j \in \{1,2,...,7 \}$, with a positive sign corresponding to an unstable mode. Note that this is not a regular eigenvalue problem since the matrix $\mathbf{A}(S_0)$ is degenerate. Nevertheless, it can be tackled with standard numerical eigensolvers, such as a generalized Schur decomposition, see, e.g., \citet{Golub1996}.

Numerical analysis of the dispersion relation confirms that the strongly magnetized \citet{Komissarov06} torus is locally stable when reduced to its non-magnetized form \citep{Abramowicz78a}. The magnetized solutions with purely toroidal fields are also stable under axisymmetric perturbations ($k_\phi = 0$). This is why the solution was found to be stable in the 2D GRMHD numerical simulations of \citet{Komissarov06}. However, this is not the case for non-axisymmetric perturbations. When we allow for $k_\phi \ne 0$, we find a single unstable mode. Observing its behavior, we find that regardless of the value of $k_r$ and $k_\theta$, the mode is azimuthal in character, i.e., $\delta u^r = \delta u^\theta = \delta b^r = \delta b^\theta = 0$ (with reasonable accuracy). Also, the growth rate is found to be proportional to $k_\phi$, while having negligible dependence on $k_r$ and $k_\theta$.

As a consistency check we verified that the unstable mode characteristics remain the same if we make a~slightly different choice for our 8 perturbed equations, i.e., it does not matter which one of 4 induction equations we replace with the orthogonality condition, equation (\ref{eq:uborthogonal}). That was to be expected, since the induction equations, in general, are not independent.

\subsection{Simplified analytic approach}

In this subsection we provide an analytic description of the simplest possible unstable non-axisymmetric perturbation mode: $k_r = k_\theta = 0$ and $k_\phi \ne 0$. For this analysis, we assume that we are looking for a~mode with $\delta u ^r = \delta u^\theta = \delta b^r = \delta b^\theta = 0$. Although these are strong assumptions, the numerical analysis of the full perturbed system (Section \ref{s.full}) indicate that they well characterize the unstable mode. We also neglect the radial structure of the background solution. Under such assumptions, the vertical Euler, vertical induction, and radial induction equations become trivial, and the azimuthal and time induction equations both reduce to a simple result
\be
\delta b^\phi = \Omega \delta b^t ~.
\ee
Now the continuity equation, together with the azimuthal and time Euler equations, form a system with 3 unknown amplitudes $(\delta u^\phi , \delta b^\phi , \delta w)$. Explicitly, after some rudimentary algebra, the simplified system reads
\be 
\rho(k_\phi + \ell \omega) \delta u^\phi + u^\phi(k_\phi + \omega/\Omega)  \delta \rho = 0 ~,
\ee
\be 
\left( h u_\phi \delta u^\phi - b_\phi \delta b^\phi \right) \left( 2 \omega + k_\phi \frac{1+\Omega \ell}{\ell}\right) -\left(\frac{\Omega k_\phi + \omega }{1-\Omega \ell} + \Gamma \frac{P_g}{w}\omega \right) \delta w = 0 ~,
\ee

\be 
\left( h u_\phi \delta u^\phi - b_\phi \delta b^\phi\right) \left( 2 k_\phi + \omega \frac{1+\Omega \ell}{\Omega}\right) -\left(\ell \frac{\Omega k_\phi + \omega }{1-\Omega \ell} - \Gamma \frac{P_g}{w}k_\phi \right) \delta w = 0 ~.
\ee
In these equations we have neglected the $g_{t \phi}$ metric component for the sake of simplicity; hence they are only quantitatively correct for a non-rotating black hole. This is clearly an inconsistency, since the background is calculated for $g_{t \phi} \neq 0$, but we accept that since we are only interested in giving an approximate formulation.

The three eigenfrequencies $\omega^{(j)}$ of the simplified system can be found easily. The first one is a real root (purely oscillatory mode)
\be
\omega^{(1)} = -k_\phi/\ell ~.
\ee
The other two, $\omega ^{(-)}$ and  $\omega^{(+)}$, are the roots of the quadratic equation
\be 
A \omega^2 + B \omega + C = 0 ~,
\ee
where
\be 
A = \ell \left[1 + \Gamma \frac{P_g}{w}(1 + \ell \Omega) \right] ~,
\ee
\be 
B = 2 \Omega \ell k_\phi \left[1 + 2\Gamma \frac{P_g}{w} \right] ~,
\ee
and
\be 
C = \Omega k_\phi^2 \left[ \Omega \ell + \Gamma \frac{P_g}{w}(1 + \ell \Omega)\right] ~.
\ee
The determinant of this equation is equal to
\be 
\Delta = -4  k_\phi^2 \Gamma \frac{P_g}{w} \ell \Omega (1-\ell \Omega)^2\left(1+\Gamma\frac{P_g}{w} \right)
\ee
which is always negative. Hence, there is {\em always} an unstable mode (root with positive imaginary part), with growth rate
\be 
\label{eq:growthrate}
\text{Im} \left \{\omega^{(+)} \right \} = \frac{\sqrt{|\Delta|}}{2A} = \frac{ k_\phi (1-\ell \Omega)\sqrt{ (\Omega/\ell) \Gamma (P_g/w)  \left(1+\Gamma P_g/w \right)}}{ \left[1 + \Gamma (P_g/w)(1 + \ell \Omega) \right]} ~.
\ee
While non-zero $b^\phi$ is crucial for this mode to exist, note that the growth rate scales with $k_\phi$ and does not depend explicitly on the magnetic field strength. Obviously, there is an implicit dependence on the magnetic pressure, since diminishing the pressure ratio parameter $\beta$ results in smaller value of $P_g$. Nevertheless, this indicate that the instability should be present for wide range of magnetic pressures, both very weak and superthermal. In the limit of low gas pressure, $P_g \rightarrow 0$, $\omega ^{(-)}$ and $\omega^{(+)}$ reduce to a single real root of value
\be 
\omega^{(0)} = - \Omega k_\phi ~,
\ee
corresponding to a co-rotation mode with zero growth rate. Hence, the mode stabilizes in the limit of zero sound speed, $c_s \rightarrow 0$, corresponding to $P_g \rightarrow 0$. This is a known property of the non-axisymmetric MRI \citep{Kim00}.

Assuming that $g_{t \phi}$ (and the black hole spin) equals zero also for the background solution, the relation between $\ell$ and $\Omega$ becomes
\be 
\ell = - \frac{g_{\phi \phi}}{g_{tt}} \Omega \equiv \mathcal{R}^2 \Omega ~,
\ee
and the unstable mode growth rate can be expressed as
\be
\text{Im} \left \{\omega^{(+)} \right \} = \frac{ k_\phi (1-\mathcal{R}^2 \Omega^2)\sqrt{\Gamma \frac{P_g}{w}  \left(1+\Gamma\frac{P_g}{w} \right)}}{\mathcal{R} \left[1 + \Gamma \frac{P_g}{w}(1 + \mathcal{R}^2 \Omega^2) \right]}~.
\label{eq:GrowthRate2}
\ee
It is interesting to observe the behavior of equation (\ref{eq:GrowthRate2}) in the limit of $\mathcal{R} \Omega \rightarrow 0$, as is the case far from the black hole, when $\Omega$ drops faster than $r^{-1}$.  An example of when this would be true is the case of a torus approaching the thin disk limit, such that $\mathcal{R} \sim r$ and $\Omega \sim r^{-3/2}$. Retaining only the dominant order of small $P_g/w$ we find
\be
\text{Im} \left \{\omega^{(+)} \right \} \approx \frac{ k_\phi}{r} \left(\Gamma \frac{P_g}{w} \right)^{1/2} \approx  \frac{k_\phi}{r} c_s .
\label{eq:GrowthRateKep}
\ee 
Once again we see that the instability vanishes in the $c_s \rightarrow 0$ limit.

\subsection{Results of the MHD system eigenanalysis}

\begin{figure}
\includegraphics[trim = 3mm 5mm 0mm 5mm, width = 0.235\textwidth]{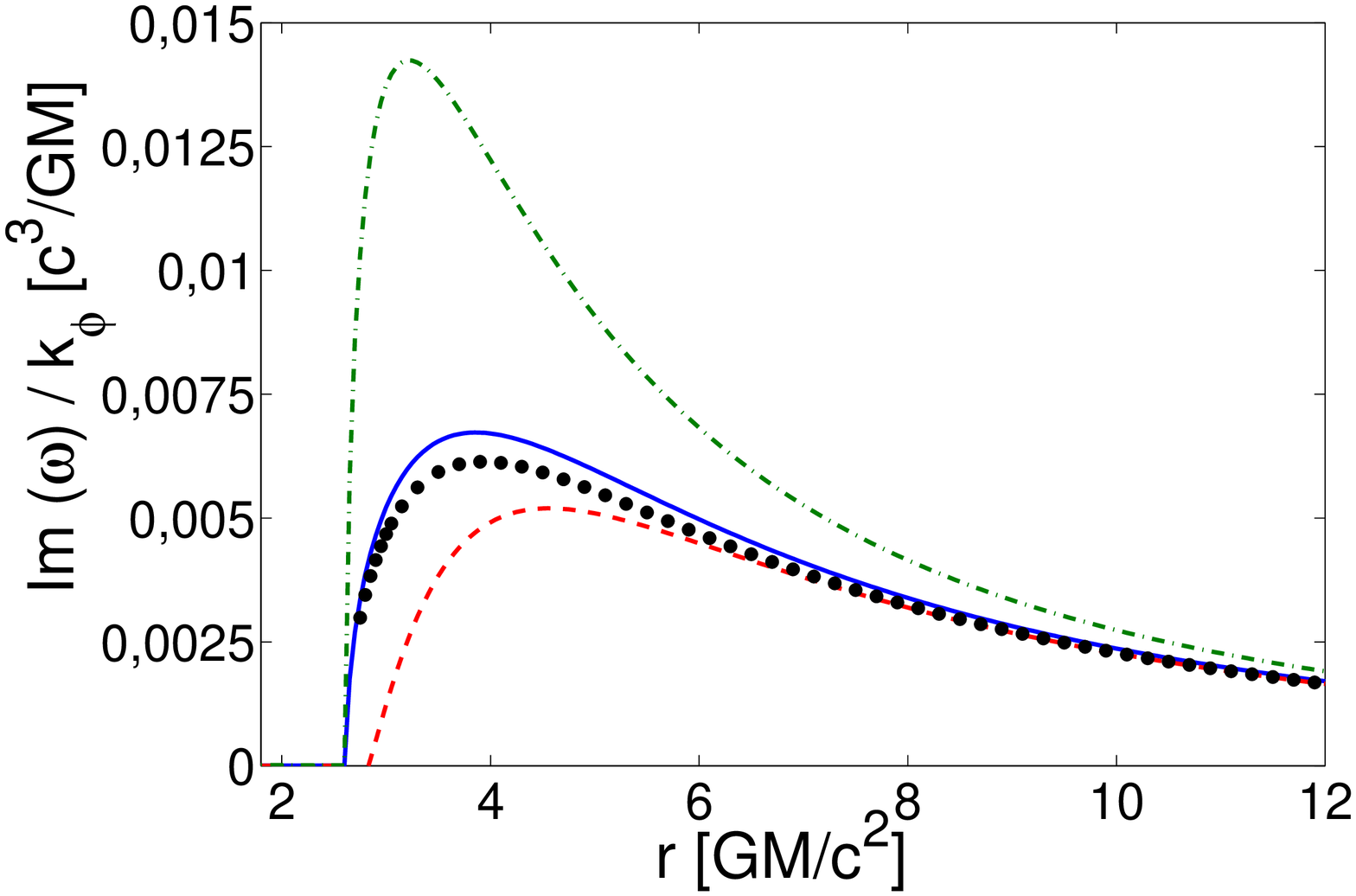}
\includegraphics[trim = 3mm 5mm 0mm 5mm, width = 0.235\textwidth]{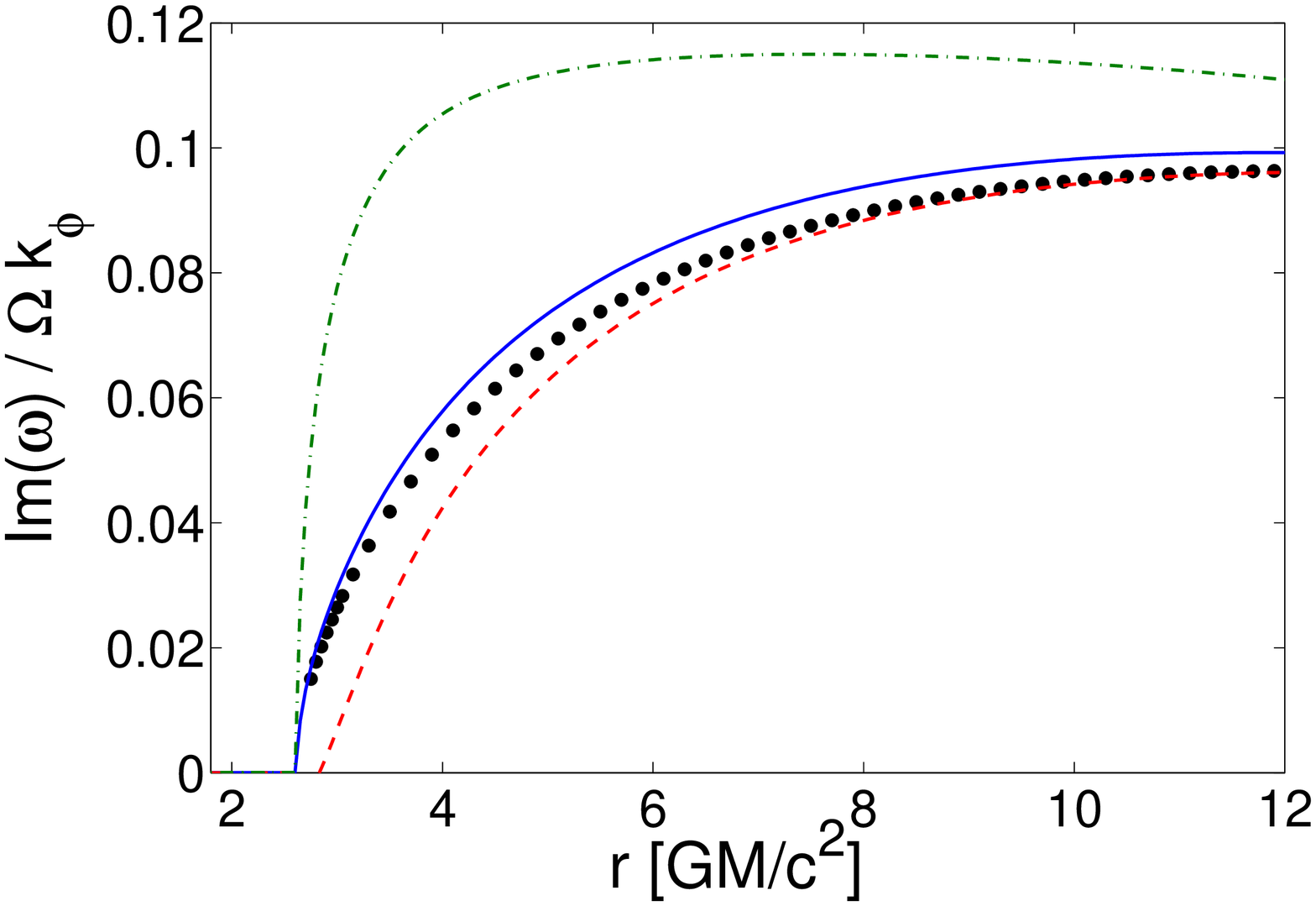}
\caption{Growth rate of the unstable mode for the case A torus as a function of radius for $k_\phi = 1$. Black dots - numerical eigenanalysis, blue continuous line - eq. (\ref{eq:growthrate}), red dashed line - eq. (\ref{eq:GrowthRate2}), green dash-dot line - eq. (\ref{eq:GrowthRateKep}). In the {\em left} panel, growth rates are given in geometric units, while in the {\em right} panel, they are given in relation to the local angular velocity $\Omega$.}
\label{fig:GRKomA}
\end{figure}

\begin{figure}
\includegraphics[trim = 3mm 5mm 0mm 5mm, width = 0.235\textwidth]{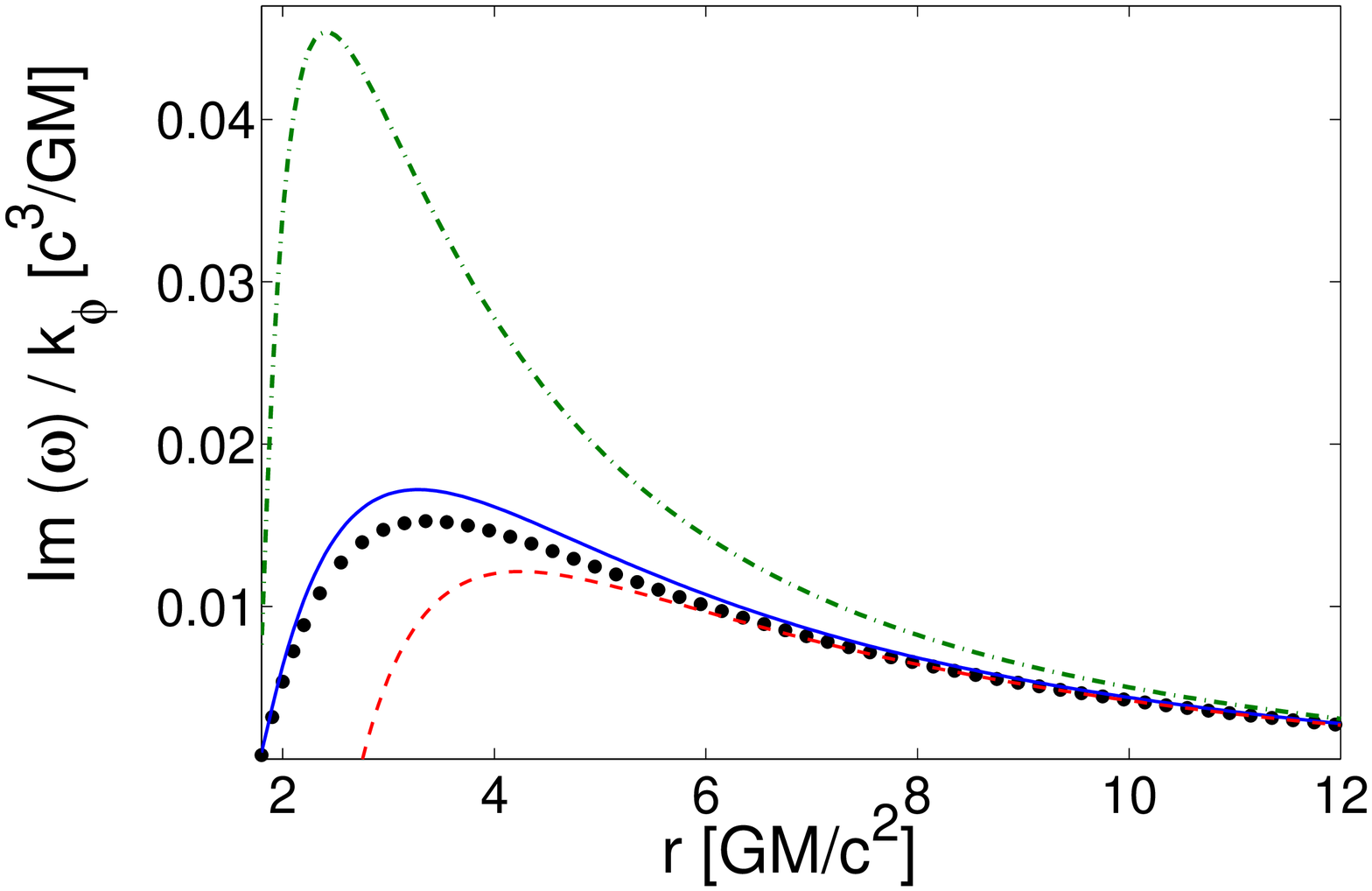}
\includegraphics[trim = 3mm 5mm 0mm 5mm, width = 0.235\textwidth]{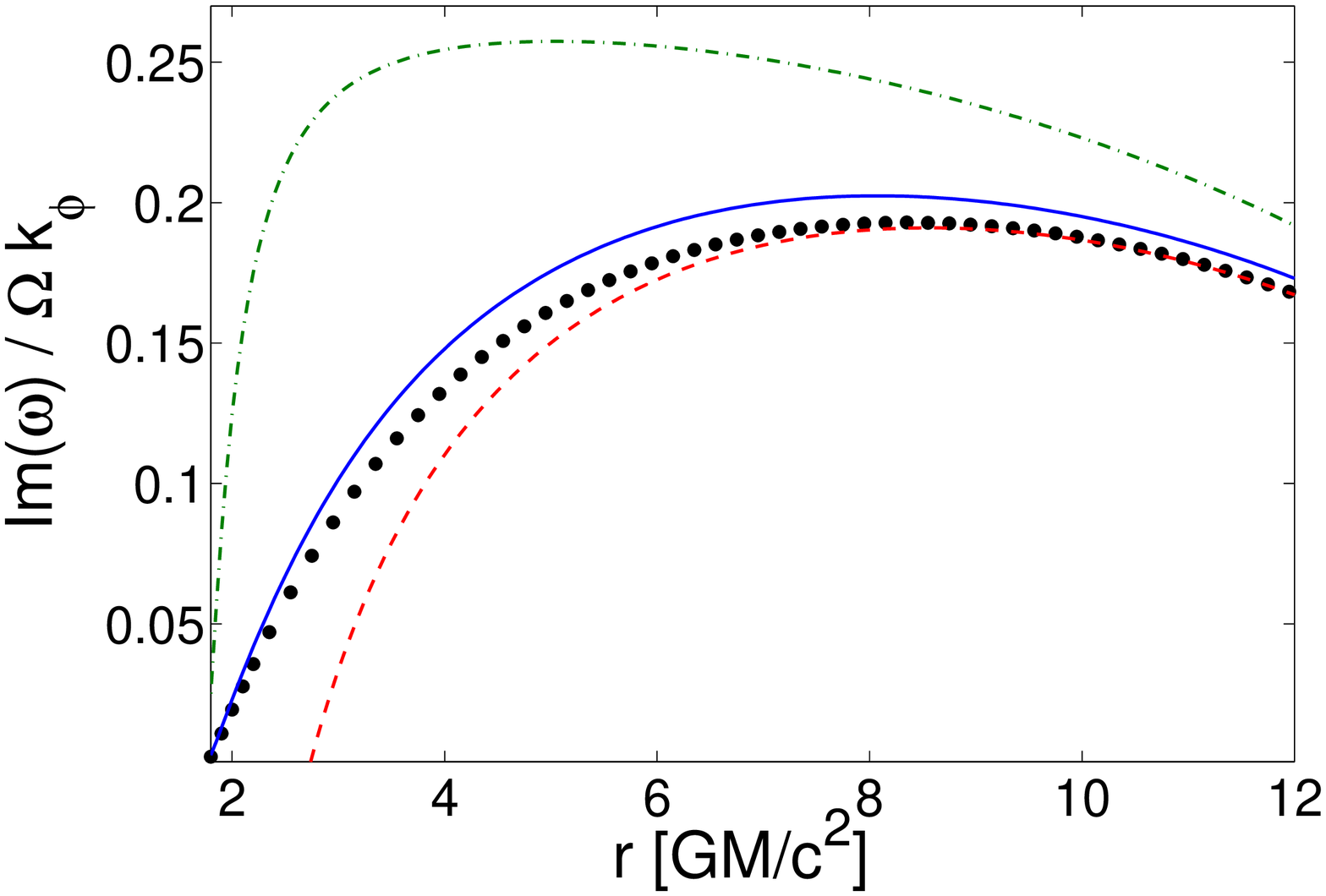}
\caption{Same as Fig. \ref{fig:GRKomA}, but for case B.}
\label{fig:GRKomB}
\end{figure}

\begin{figure}
\includegraphics[trim = 3mm 5mm 0mm 5mm, width = 0.235\textwidth]{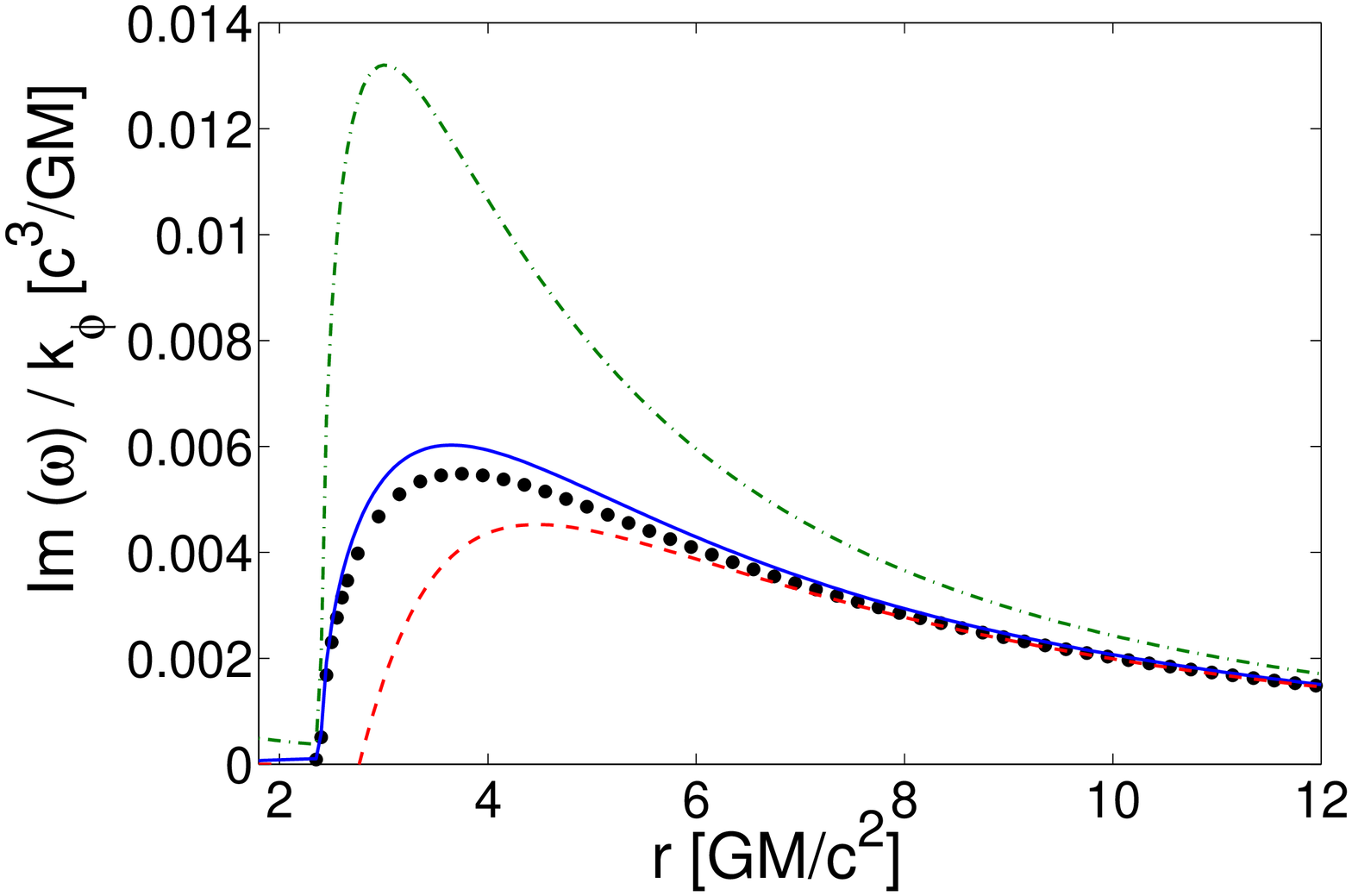}
\includegraphics[trim = 3mm 5mm 0mm 5mm, width = 0.235\textwidth]{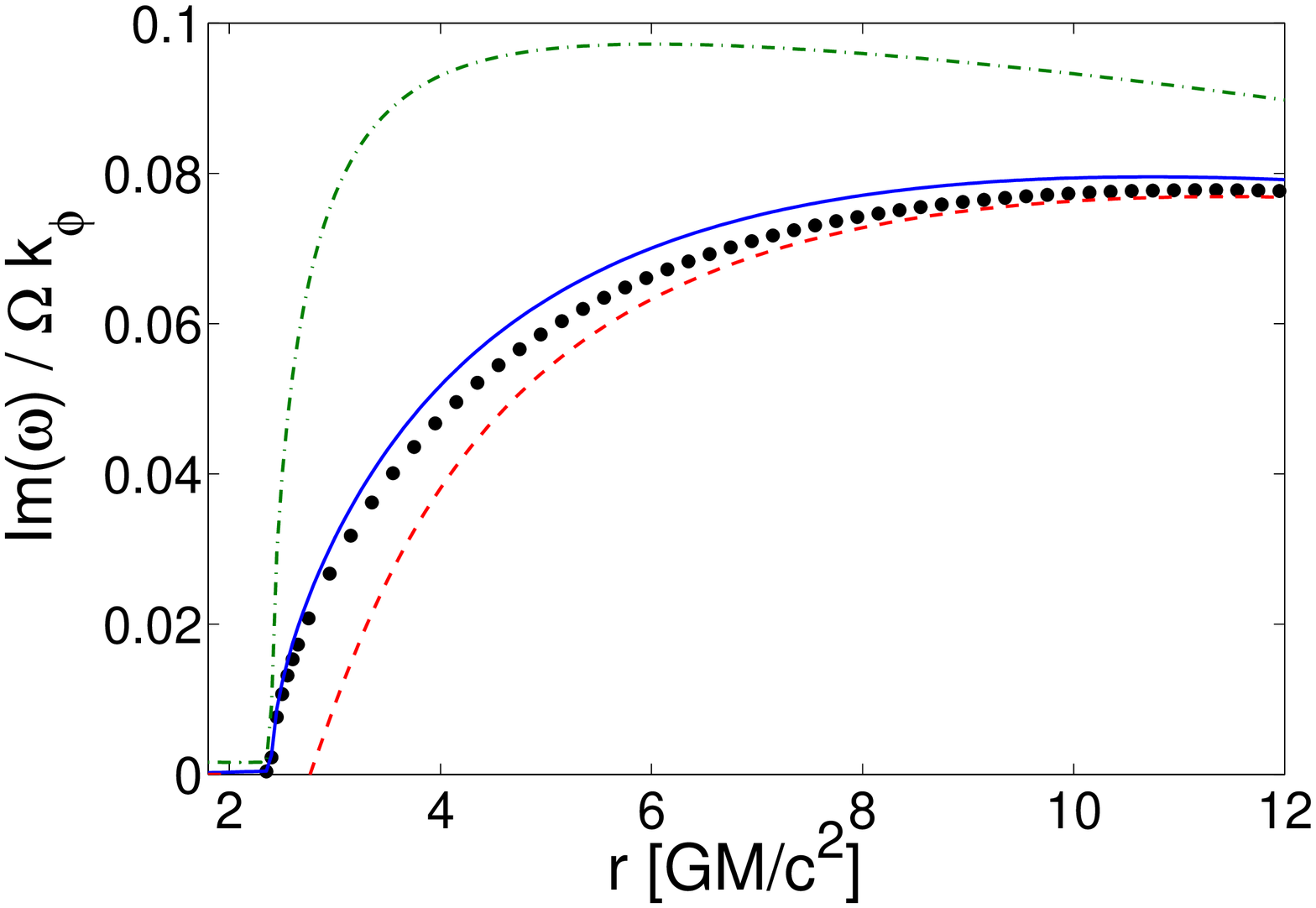}
\caption{Same as Fig. \ref{fig:GRKomA}, but for case C.}
\label{fig:GRKomC}
\end{figure}

In Figs. \ref{fig:GRKomA}-\ref{fig:GRKomC}, we present the radial dependancies of the unstable mode growth rates for the three torus configurations considered. The {\em left} plot compares the results of numerical eigenanalysis on the full system of the perturbed MHD equations with the analytic results given by equations (\ref{eq:growthrate}), (\ref{eq:GrowthRate2}), and (\ref{eq:GrowthRateKep}). The {\em right} plot presents the same results, but in units of the local fluid angular velocity $\Omega$. Since the growth rates scale linearly with the azimuthal wavenumber, $k_\phi$, as indicated by equation (\ref{eq:growthrate}) (and confirmed by the numerical eigenanalysis), all results have been normalized to $k_\phi = 1$.


While the growth rate does not explicitly depend on the pressure-ratio parameter $\beta$, we observe that the growth rate is smaller whenever the magnetic pressure is more dominant (i.e. smaller $\beta$).  For example, case A has a smaller growth rate than case B. This is because a smaller $\beta$ requires a lower gas pressure $P_g$, and the growth rate is roughly proportional to $P_g^{1/2}$. Conversely, we observe little influence of the angular momentum variability on the growth rate; for instance, case A and case C exhibit very similar rates. Finally, one may conclude that the analytic formula (\ref{eq:growthrate}) captures sufficiently well the behavior of the mode found in the numerical eigenanalysis. However, neglecting $\mathcal{R} \Omega$ in the growth rate, as it was done in equation (\ref{eq:GrowthRateKep}) and shown with green dash-dot lines in Figs. \ref{fig:GRKomA}-\ref{fig:GRKomC}, does not yield quantitatively correct results in the relativistic regime.

\subsection{Shearing background influence}
\label{sect:shearingBckg}
In the current calculations, the influence of the shearing background \citep{Goldreich65} is not taken into account. It is not straightforward to include this effect consistently in the relativistic analysis. If it had been taken into account, its effect would be to generate a growing-in-time radial wavenumber $k_r$, which has a stabilizing influence on the system. In our testing, we established that a non-zero $k_r$ does not change the unstable mode behavior in our necessarily local in time analysis. However, a non-zero time derivative of $k_r$ may. Qualitatively this may not matter in our work, as \citet{Balbus92} indicate that the shearing effect is most pronounced for larger azimuthal wave numbers, whereas we only discuss results for rather small wave numbers $k_\phi = 4$ and 8. Nevertheless, we expect the eigenanalysis results quality to deteriorate with lower ratio of growth rate to $r \text{d} \Omega / \text{d} r$, latter being the characteristic frequency for the background shearing.

\section{Global GRMHD simulations}
\label{s.numeric}

Having set up and analyzed the stability of the \citet{Komissarov06} magnetized torus solutions, we now use the {\em Cosmos++} computational astrophysics code \citep{Anninos05,Fragile12} to numerically evolve both two- and three-dimensional versions of such tori.  In 2D, this is done on a $252 \times 256$ grid, with cells spaced evenly in the coordinates $x_1$ and $x_2$.  $x_1$ is related to the normal radial coordinate through the logarithmic transformation, $x_1 \equiv 1 + \ln(r/r_\mathrm{BH})$, where $r_\mathrm{BH} = (1 + \sqrt{1-a^2})$ is the black hole radius.  $x_2$ is related to $\theta$ as $\theta(x_2)=\pi/2 [1+(1-\varepsilon)(2x_2-1)+\varepsilon(2x_2-1)^n]$, where we take $\varepsilon = 0.7$ and $n = 29$ \citep{Noble10}.  Outflow boundaries are used at $r_\mathrm{min} = 0.95 r_\mathrm{BH}$ and $r_\mathrm{max} = 50$, while reflecting boundaries are employed at $\theta = 0$ and $\pi$.

For the 3D simulations, the initial torus solution is mapped to all azimuthal zones.  We evolve only a quarter of the azimuthal domain, $0 \le \phi \le \pi/2$, with periodic boundaries at $\phi = 0$ and $\phi = \pi/2$, with a resolution of 64 zones.  For this reason, we are only able to study azimuthal modes with mode numbers, $k_\phi$, that are integer multiples of 4.  We consider all 3 cases from Table \ref{tab:params}.  We use the orbital period at $r_{c}$, i.e., $t_\mathrm{orb} = 2\pi/\Omega(r_{c})$, as a convenient unit of time; we run each simulation for $\approx 4 t_\mathrm{orb}$.

\subsection{Results of 2D Simulations}

\begin{figure}
\includegraphics[width = 0.4\textwidth]{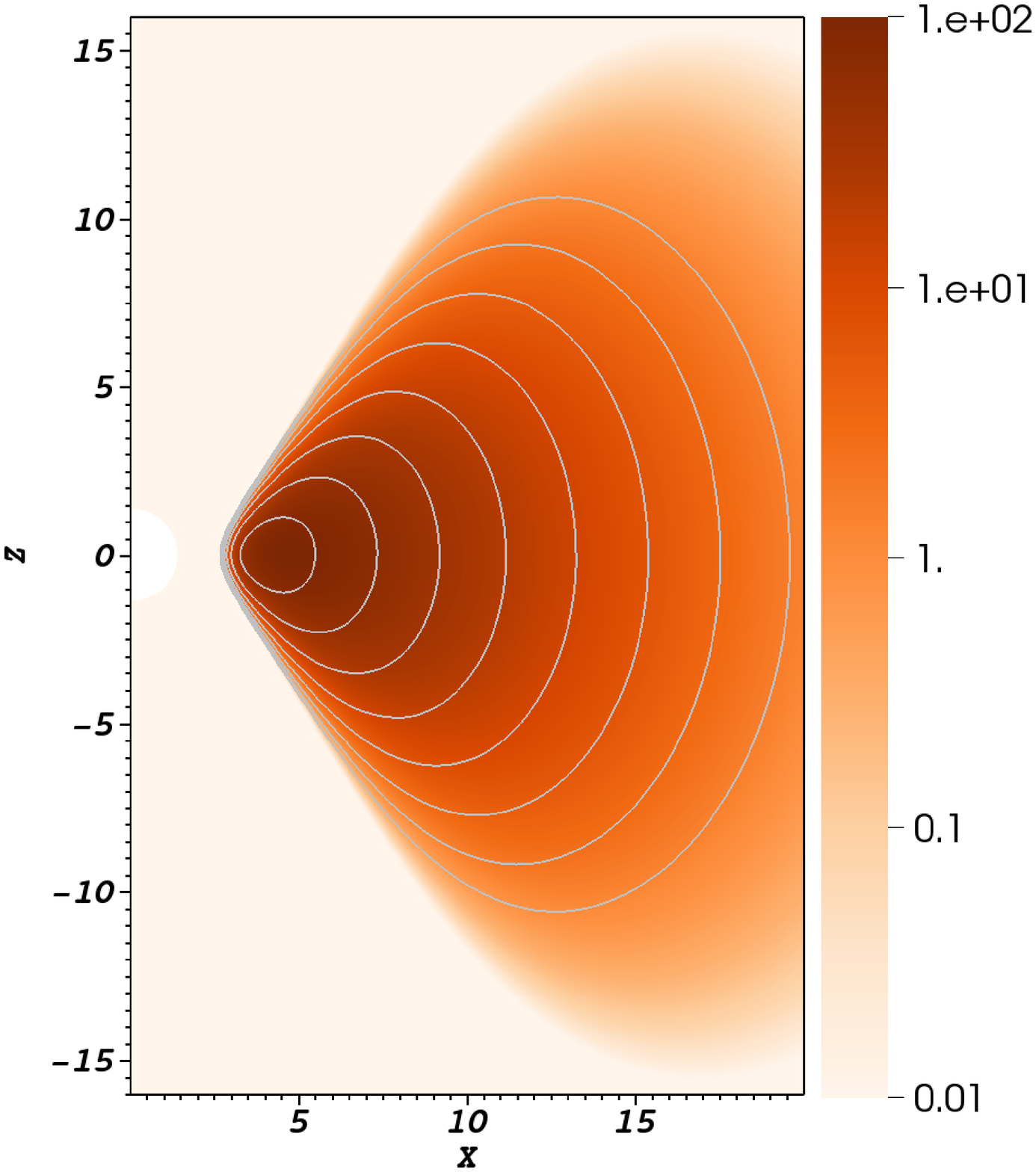}
\includegraphics[width = 0.4\textwidth]{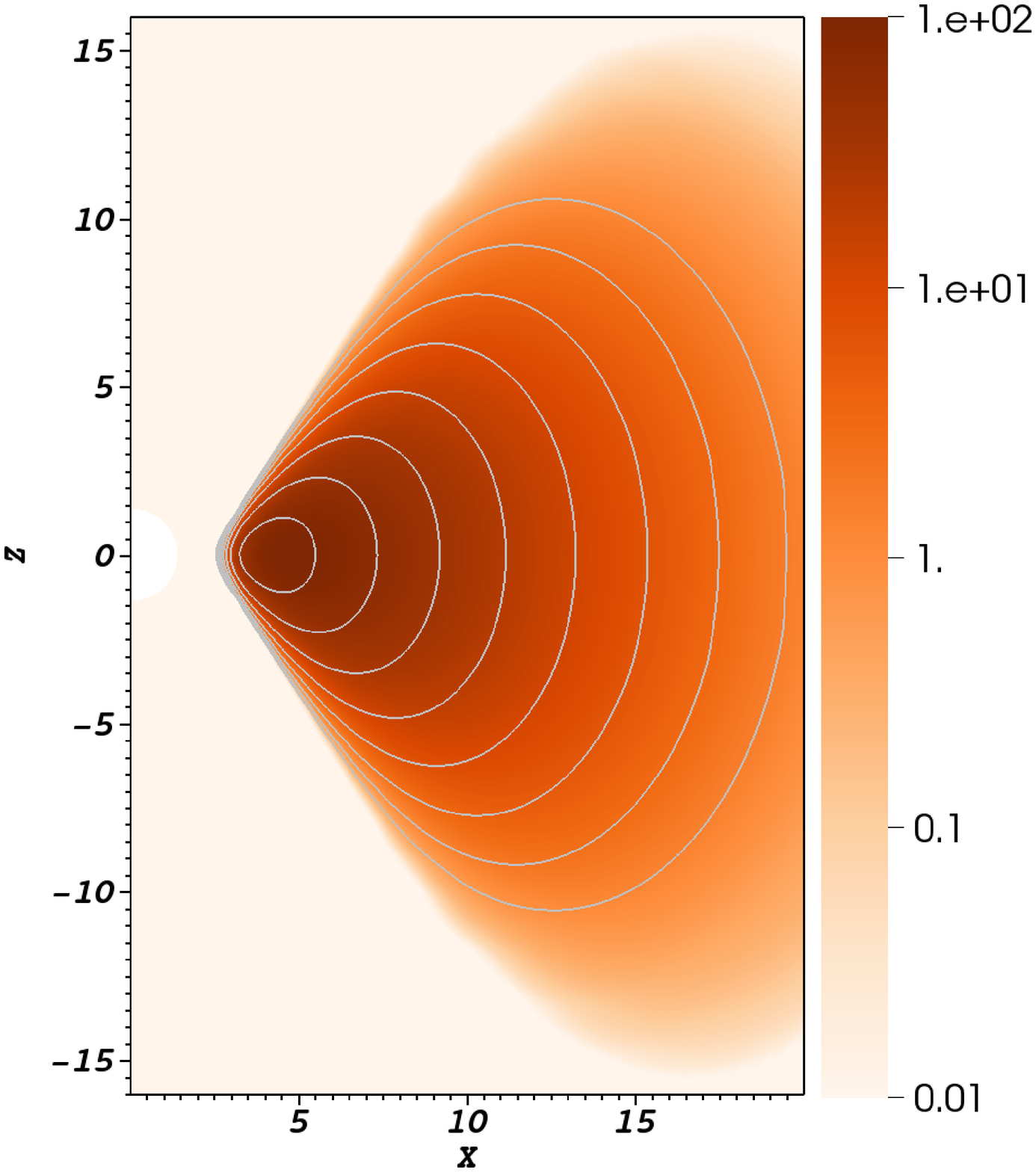}
\caption{Comparison of the initial and final states of the 2D simulation of case A.  Pseudocolor represents the gas density, $\rho$, while the contours correspond to magnetic pressure, $P_m$, and are spaced logarithmically. Results of this simulation in a~movie format are available as supplementary online material.}
\label{fig:2D}
\end{figure}

In two dimensions, the strongly magnetized torus solution of \citet{Komissarov06} is stable, as was shown in that paper.  We confirm this, as portrayed in Fig. \ref{fig:2D}, which shows the initial and final states of our 2D simulation for case A.  Aside from some small ripples along the surface of the torus, the distribution of gas and magnetic field is nearly unchanged after $4 t_\mathrm{orb}$.  Similar results are found for cases B and C. This confirms what our perturbative analysis predicts, i.e. that there are no unstable axisymmetric modes.

\subsection{Results of 3D Simulations}

\begin{figure}
\includegraphics[width = 0.4\textwidth]{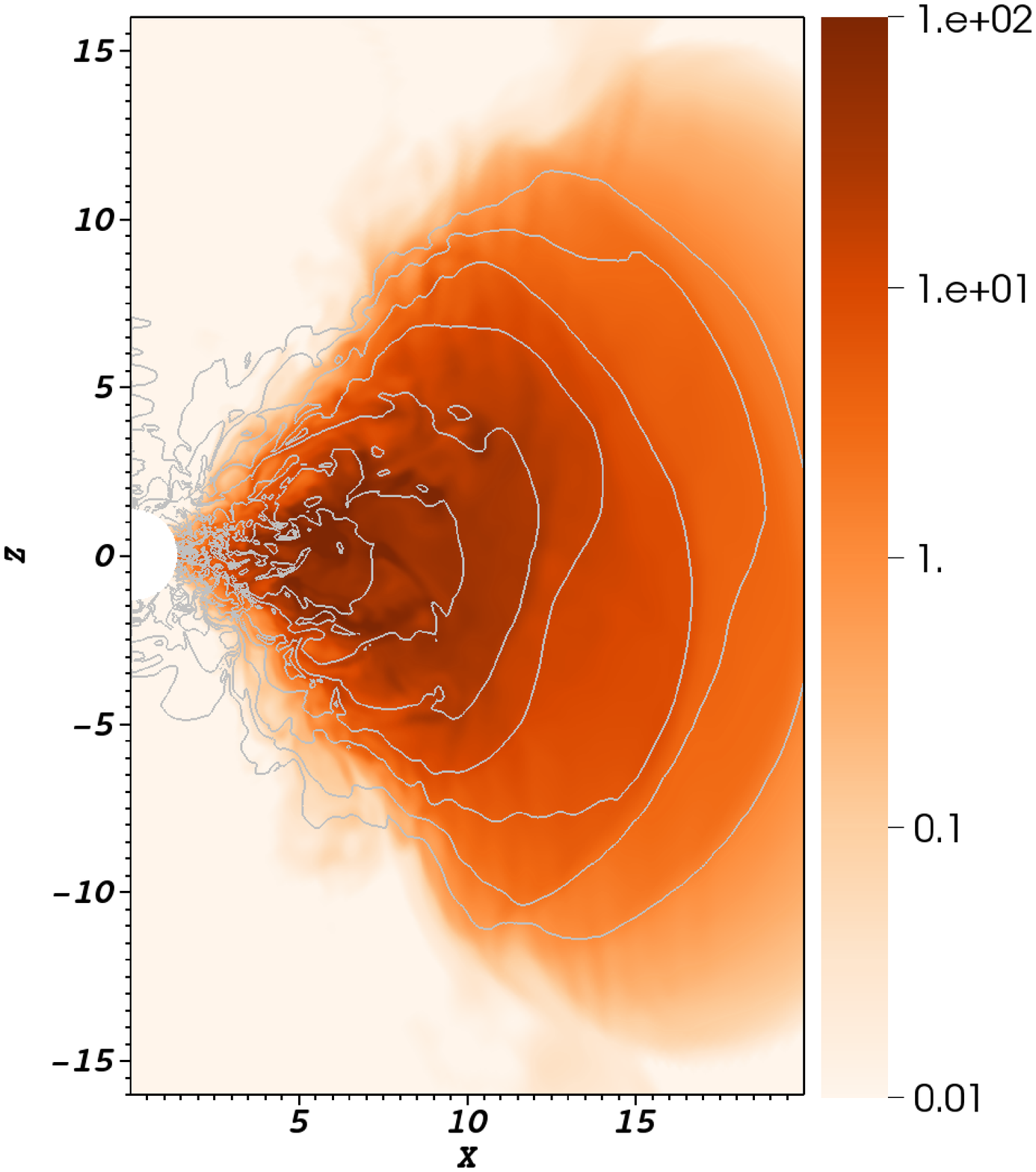}
\includegraphics[width = 0.5\textwidth]{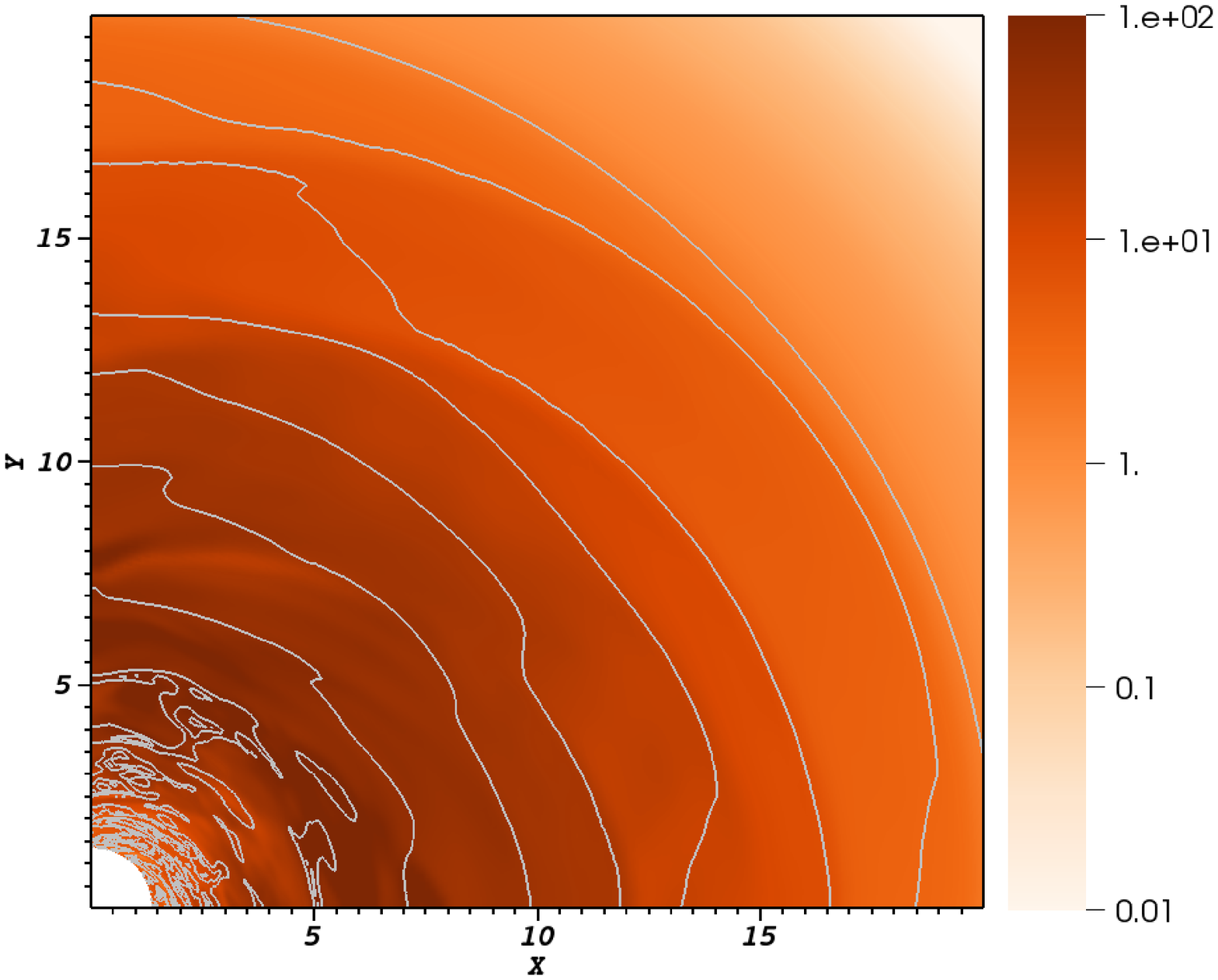}
\caption{Poloidal and midplane slices of the final state of the 3D simulation of case A. The color scale and contours are the same as in Fig. \ref{fig:2D}. Results of this simulation in a~movie format are available as supplementary online material.}
\label{fig:3D}
\end{figure}

Figure \ref{fig:3D} shows the results for our 3D simulation of case A.  The differences from Fig. \ref{fig:2D} are obvious.  Both the density and magnetic field are highly disturbed, especially close to the black hole.  What appear to be turbulent eddies are apparent on multiple scales.  The inner boundary of the torus has been completely erased, with significant gas density seen all the way down to the event horizon, i.e., we observe a~turbulence-triggered inward mass transport. The midplane image also shows extended azimuthal structures in both the density and magnetic pressure on spatial scale corresponding to rather low azimuthal wavenumbers $k_\phi$.

\section{Results comparison}

Clearly, some sort of non-axisymmetric instability was triggered in our 3D numerical simulations, but this does not guarantee that it is the same instability as predicted by our perturbative analysis.  To assess whether it may, in fact, be the same, we can compare the predicted growth rates and oscillatory frequencies with what is found in the simulations. We perform this comparison at $r = 3.5M$, since this radius gives relatively large predicted growth rates and sufficiently many orbital periods of simulation run time.

\subsection{Growth rates comparison}

\begin{figure}
\includegraphics[trim = 5mm 0mm 6mm 3mm, width = 0.24\textwidth]{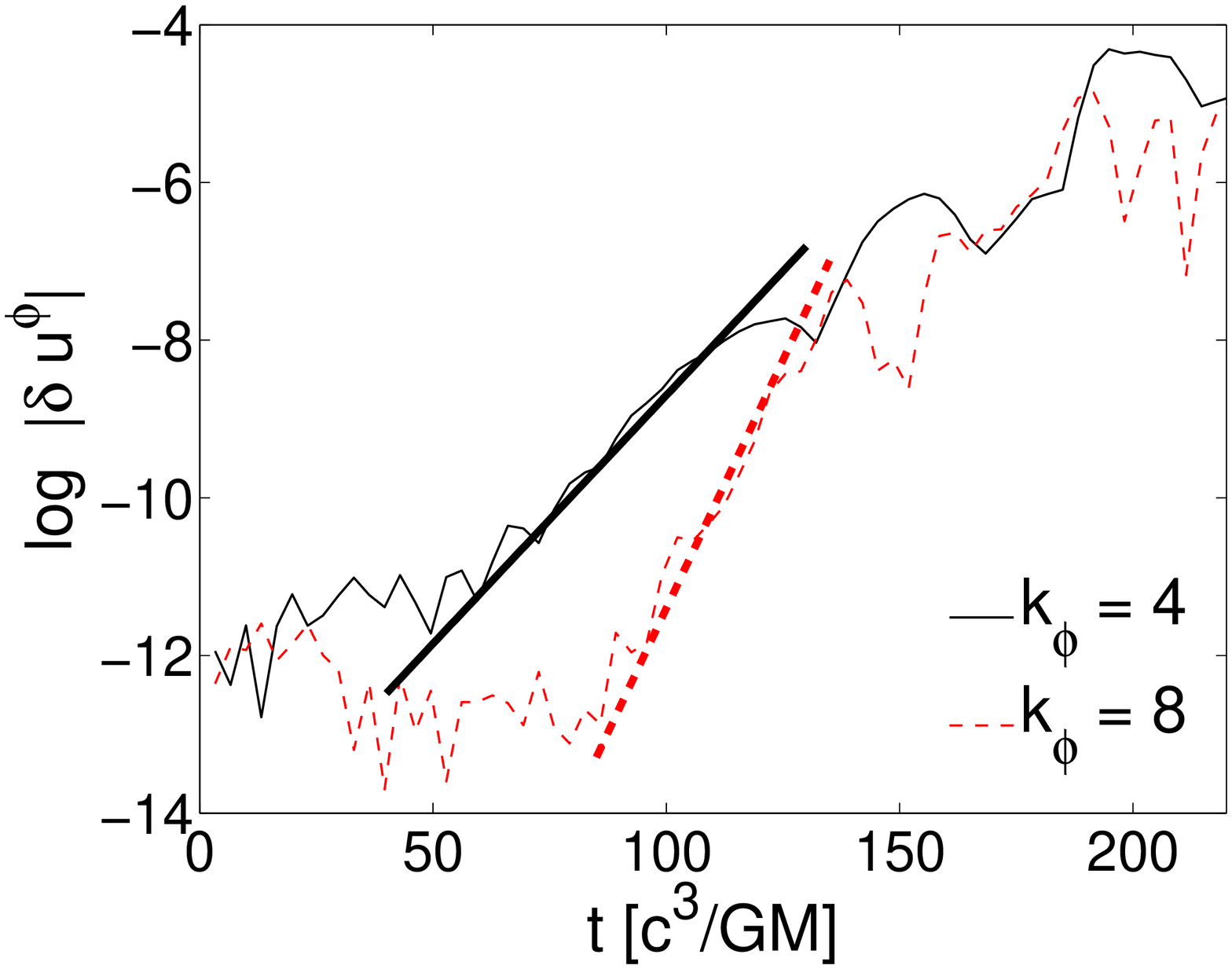}
\includegraphics[trim = 5mm 0mm 6mm 3mm, width = 0.24\textwidth]{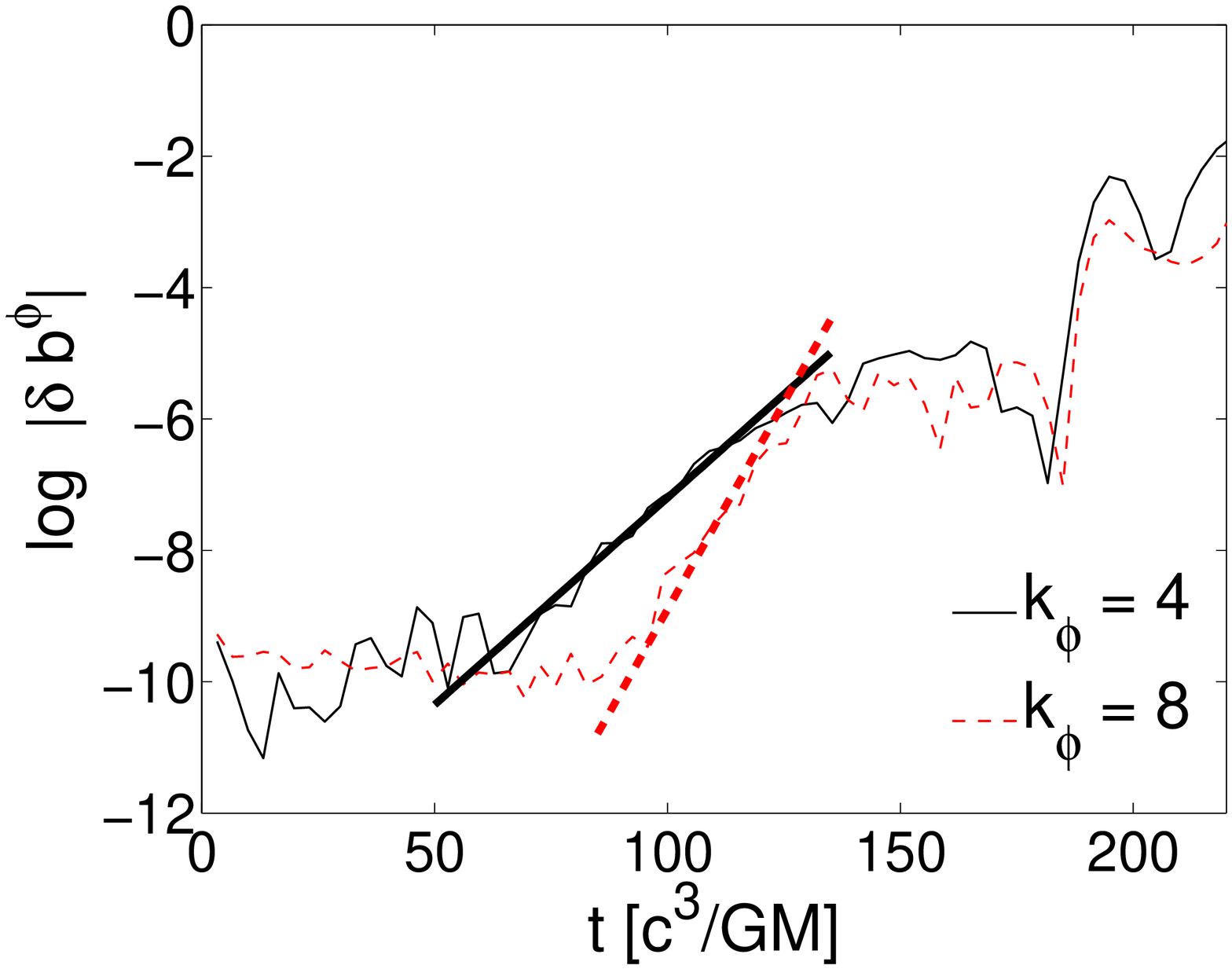}
\includegraphics[trim = 5mm 0mm 6mm 3mm,,width = 0.24\textwidth]{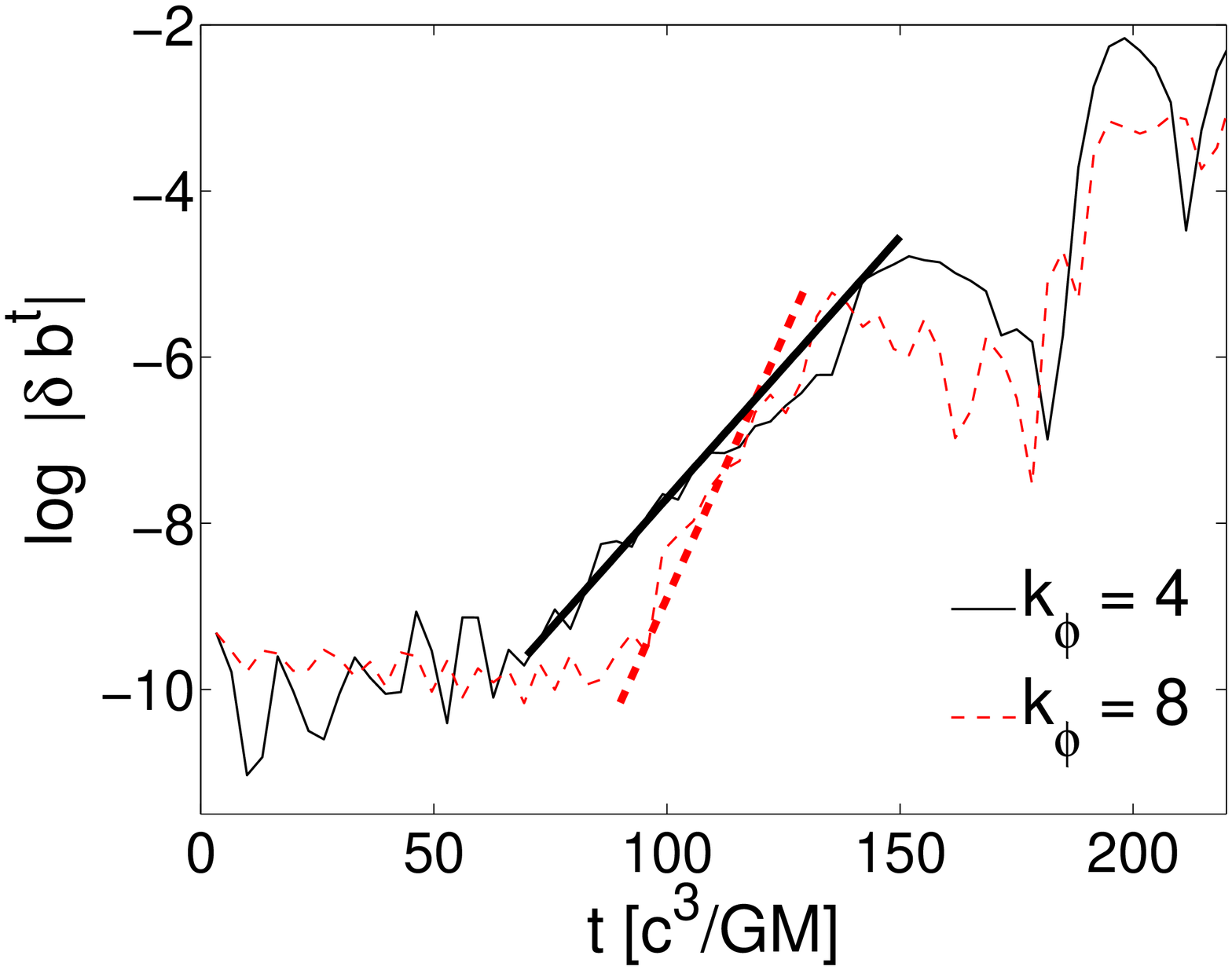}
\includegraphics[trim = 5mm 0mm 6mm 3mm, width = 0.24\textwidth]{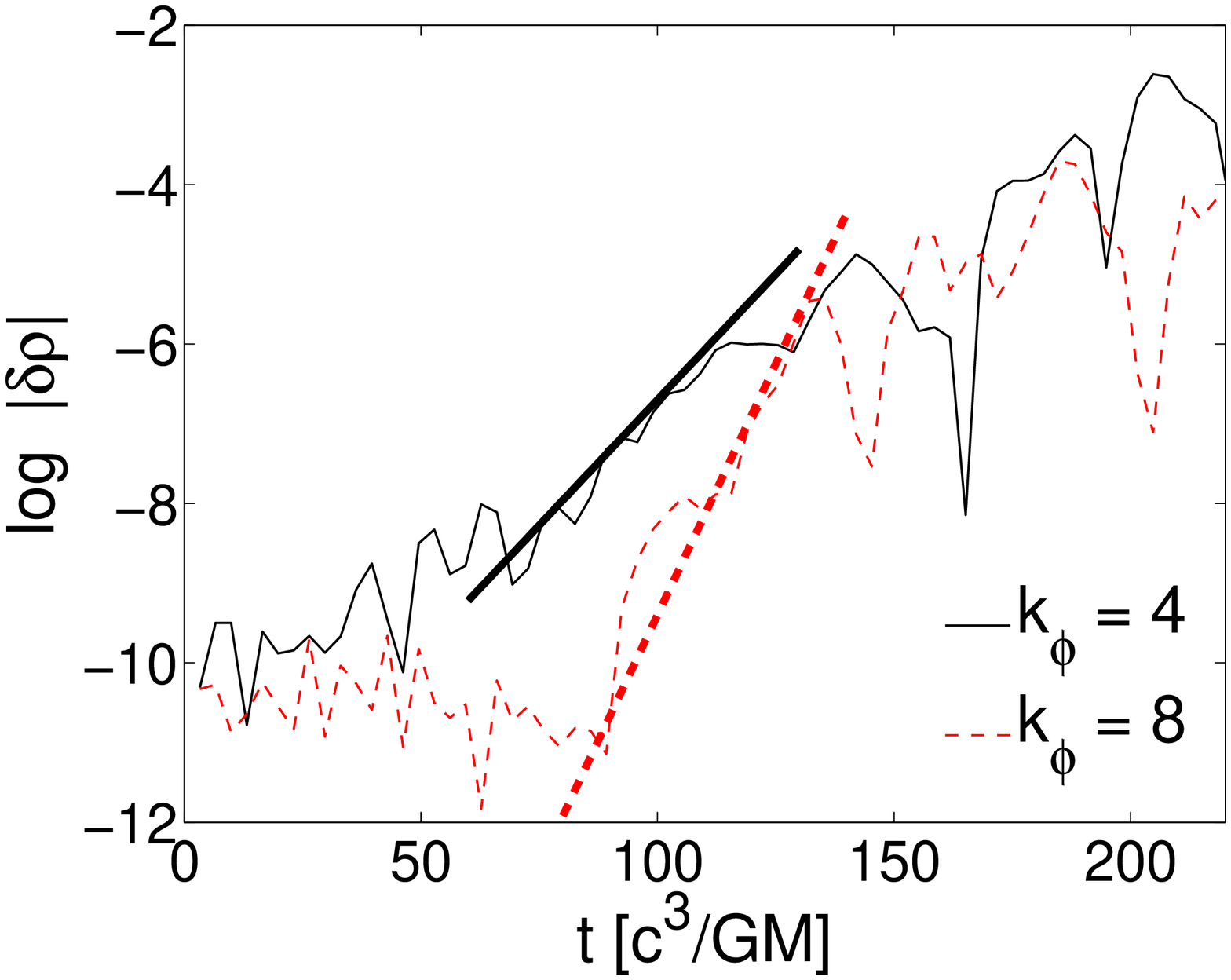}
\caption{Growth of the $k_\phi=4$ and $8$ Fourier mode amplitudes as seen in the case B simulation for the unstable variables $u^\phi$,  $b^\phi$, $b^t$ and $\rho$.  The {\em bold} lines provide the predicted growth rates (with arbitrary normalizations), based upon the linear perturbative analysis.}
\label{fig:modesB}
\end{figure}

\begin{figure}
\includegraphics[trim = 3mm 0mm 6mm 3mm, width = 0.242\textwidth]{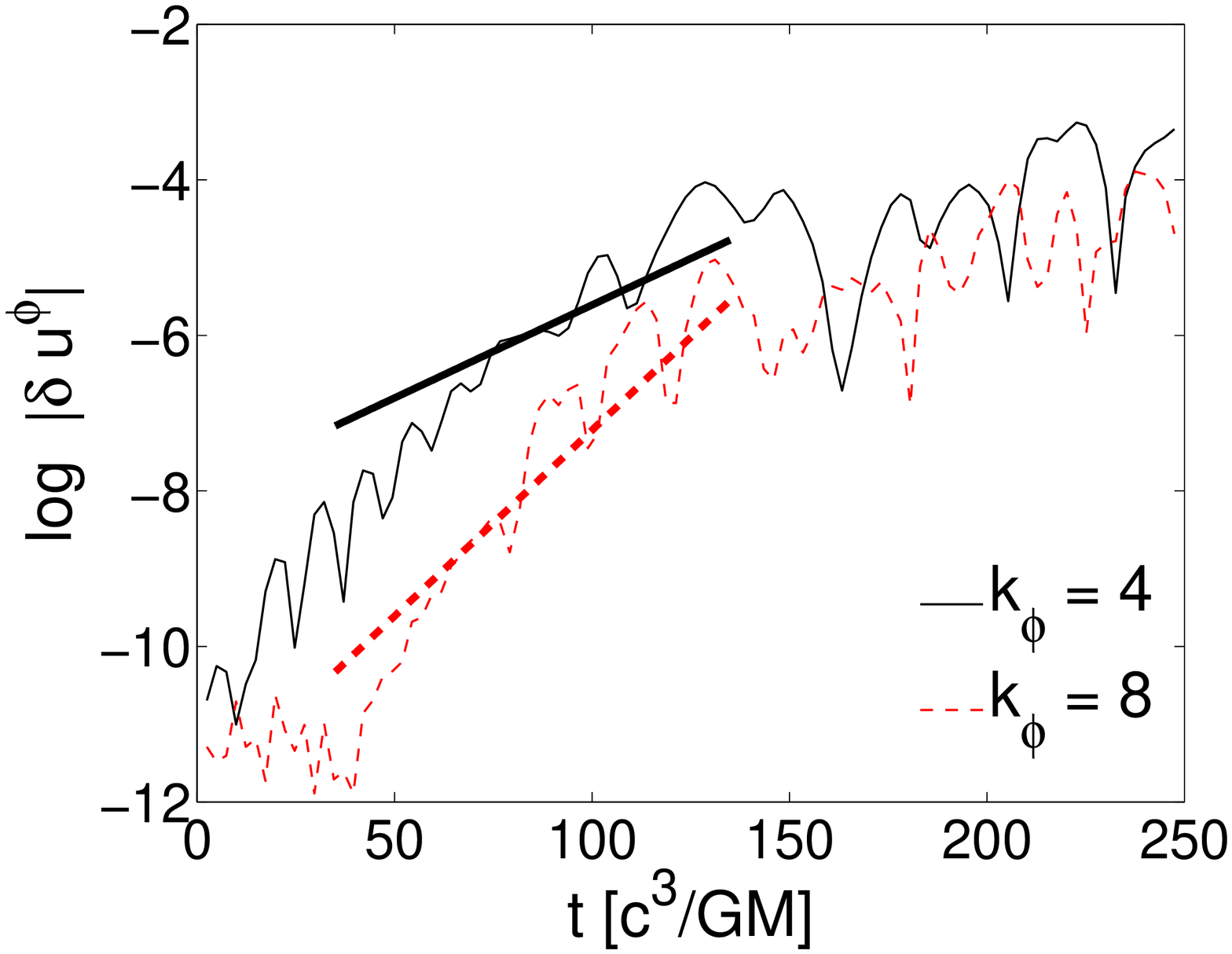}
\includegraphics[trim = 3mm 0mm 6mm 3mm, width = 0.242\textwidth]{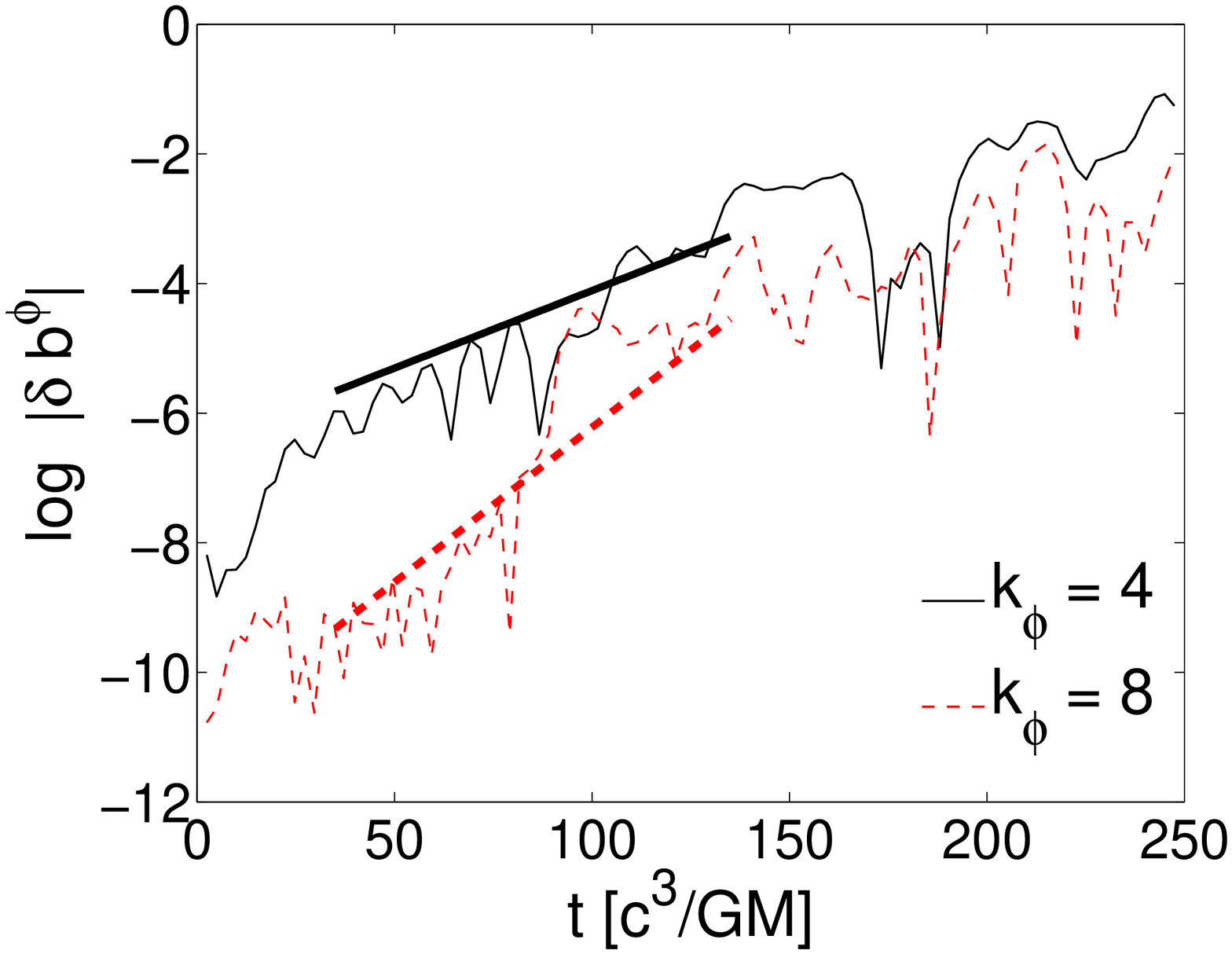}
\includegraphics[trim = 3mm 0mm 6mm 3mm, width = 0.242\textwidth]{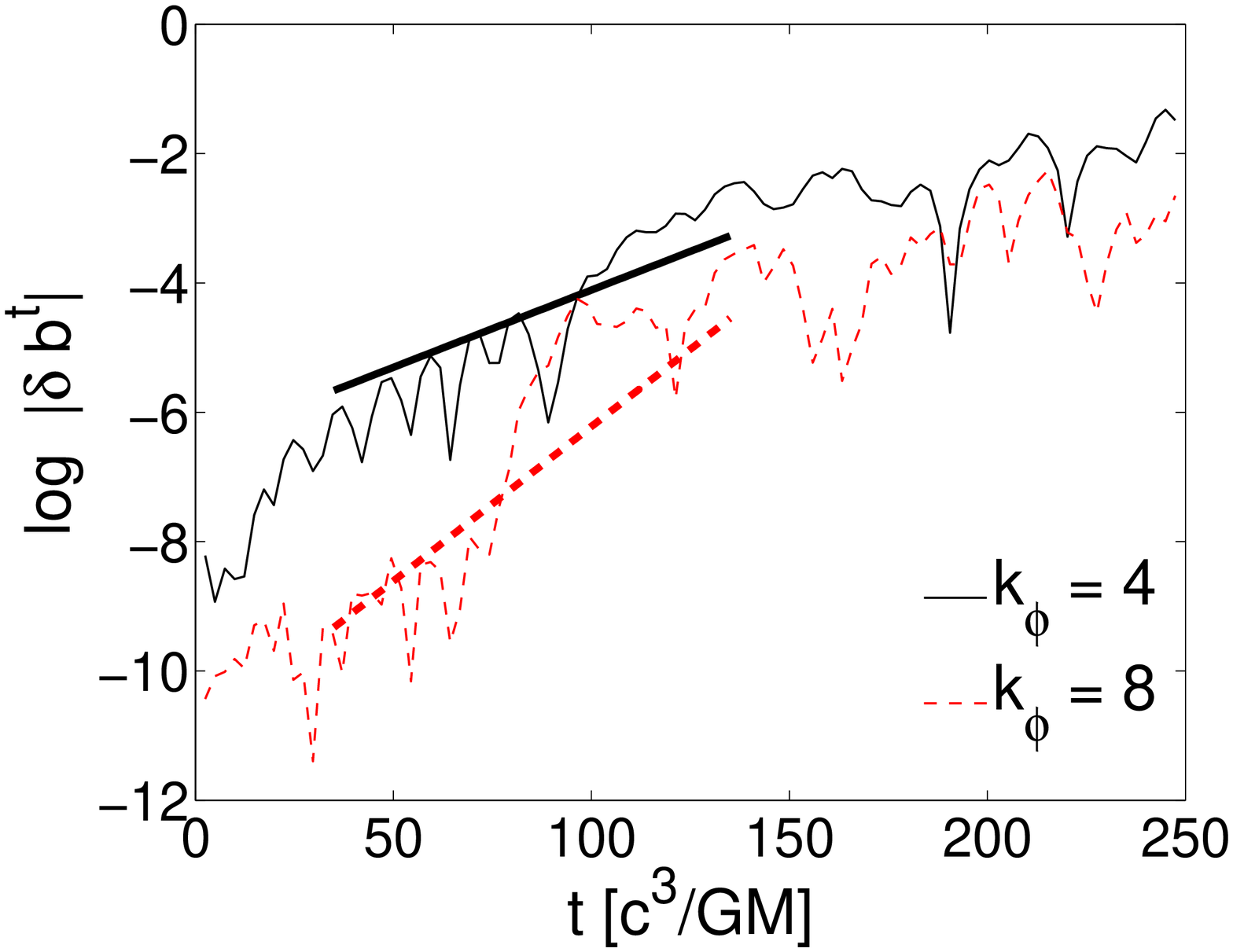}
\includegraphics[trim = 3mm 0mm 6mm 3mm, width = 0.242\textwidth]{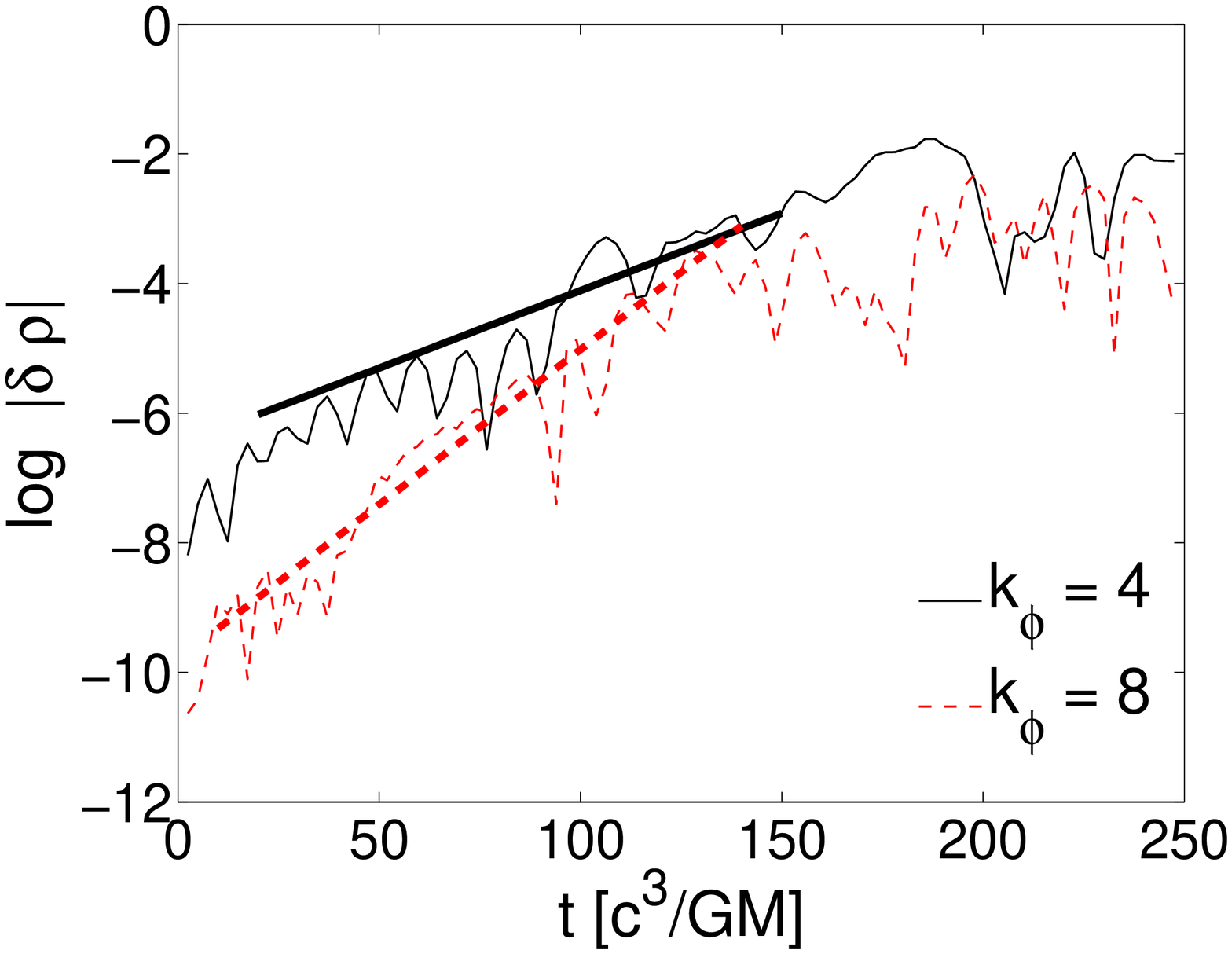}
\caption{Same as Fig. \ref{fig:modesB} except for case A.}
\label{fig:modesA}
\end{figure}

\begin{figure}
\includegraphics[trim = 3mm 0mm 6mm 3mm,, width = 0.242\textwidth]{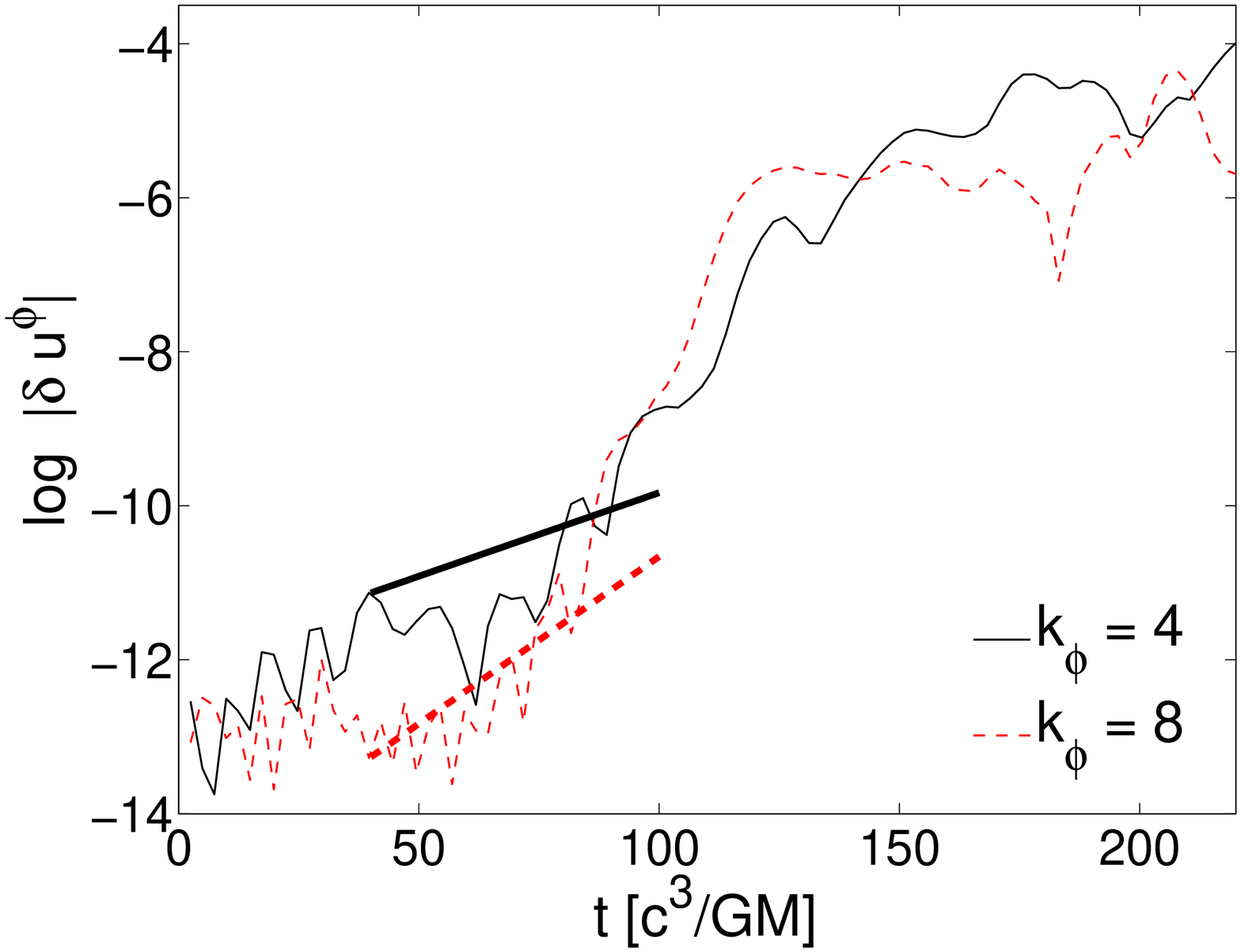}
\includegraphics[trim = 3mm 0mm 6mm 3mm, width = 0.242\textwidth]{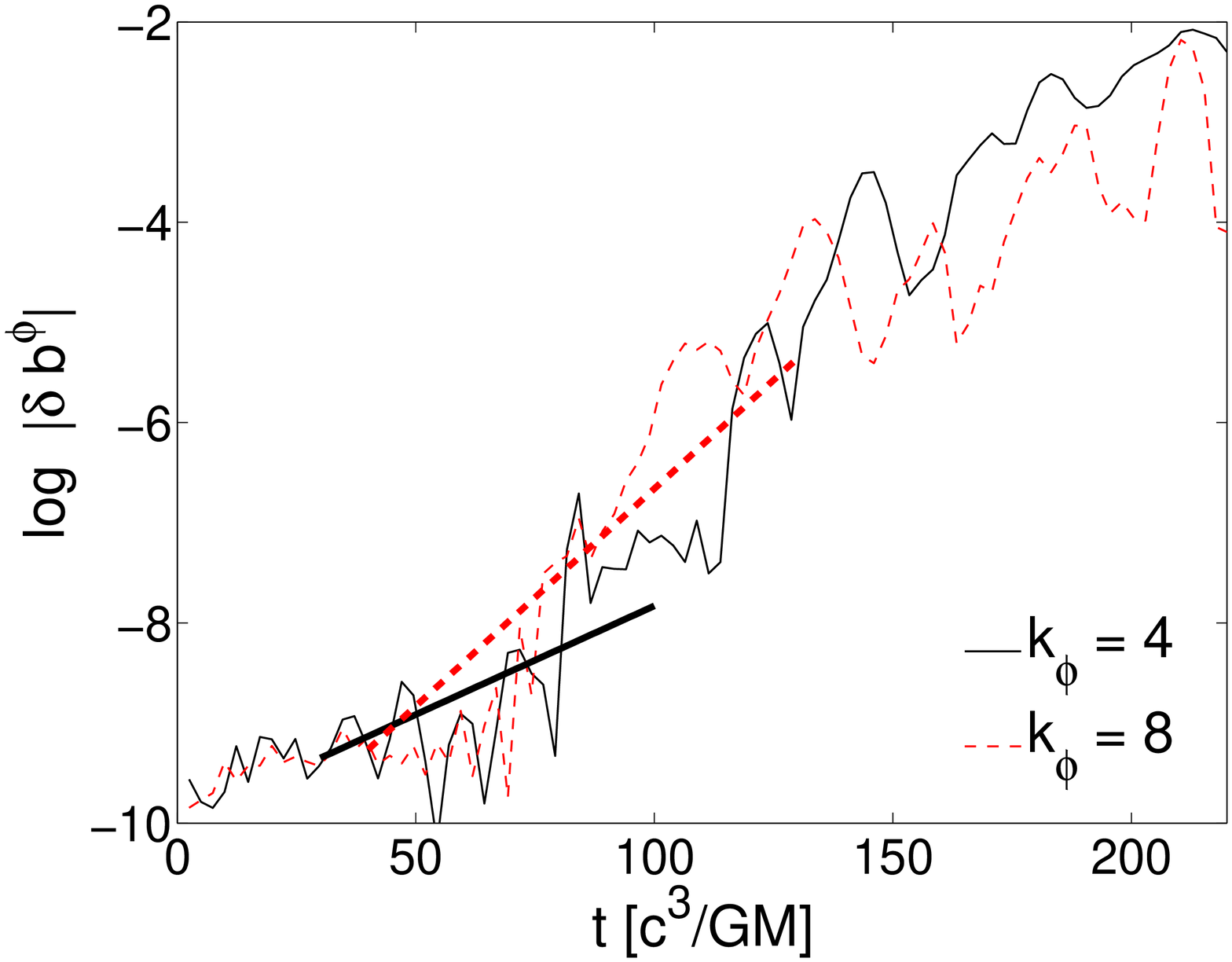}
\caption{Same as Fig. \ref{fig:modesB} except for case C and the unstable variables $u^\phi$ and $b^\phi$.}
\label{fig:modesC}
\end{figure}

Qualitatively, one can immediately see that the instability in the GRMHD simulations develops most rapidly in the inner regions of the torus, as expected from the linear analysis (Figs. \ref{fig:GRKomA}-\ref{fig:GRKomC}).  For a~more quantitative comparison, in Figs. \ref{fig:modesB}-\ref{fig:modesC}, we plot the time behavior of the azimuthal mode amplitudes. The mode amplitudes are calculated from the GRMHD simulations as $ \vert \delta n(k_\phi) \vert = f(k_\phi)/f(0)$, where
\begin{equation}
f(k_\phi) = \left|  \frac{2}{\pi} \int_0^{\pi/2} n \exp(- \mbox{i} k_\phi \phi) \mbox{d} \phi  \right|~.
\label{eq:fabs}
\end{equation}
The integral in equation (\ref{eq:fabs}) is carried out over the restricted ranges $3.375M \le r \le 3.625M$ and $\pi/2 - 0.05 \le \theta \le \pi/2 + 0.05$. Straight lines, corresponding to exponential growth with rates predicted by the eigenanalysis performed in Section \ref{s.perturb}, are overlaid on Figs. \ref{fig:GRKomA}-\ref{fig:GRKomC}.

We generally find that the mode growth rates observed in the numerical simulation match reasonably well with the predictions of the perturbation study. The agreement is particularly good for case B (Fig. \ref{fig:modesB}), for which there is a clear division of the simulation into three stages: a quasi-steady state ($0 \lesssim t \lesssim 75$, geometric units); a linear growth phase ($75 \lesssim t \lesssim 150$); and a saturated, nonlinear perturbation state ($t \gtrsim 150$). A similar correspondence of rates is observed for case A, as shown in Fig. \ref{fig:modesA}.  However, the correspondence is rather poor for the non-constant angular momentum case C, as shown in Fig. \ref{fig:modesC}. The difference may be due to our neglecting the background shear. For instance, the growth rates in case B (0.063 and 0.126 in geometric units) are almost 3 times larger than those in case C (0.022 and 0.044). Hence, the modes are more likely to appear in the simulation before being sheared out by the growing $k_r$ wave number (the characteristic rate of shearing corresponds to $r \mbox{d} \Omega / \mbox{d} r$, which differs by less than 10\% between the three considered cases). 

\subsection{Fitting the oscillation frequency}

\begin{figure}
\includegraphics[trim = 5mm 0mm 6mm 5mm, width = 0.242\textwidth]{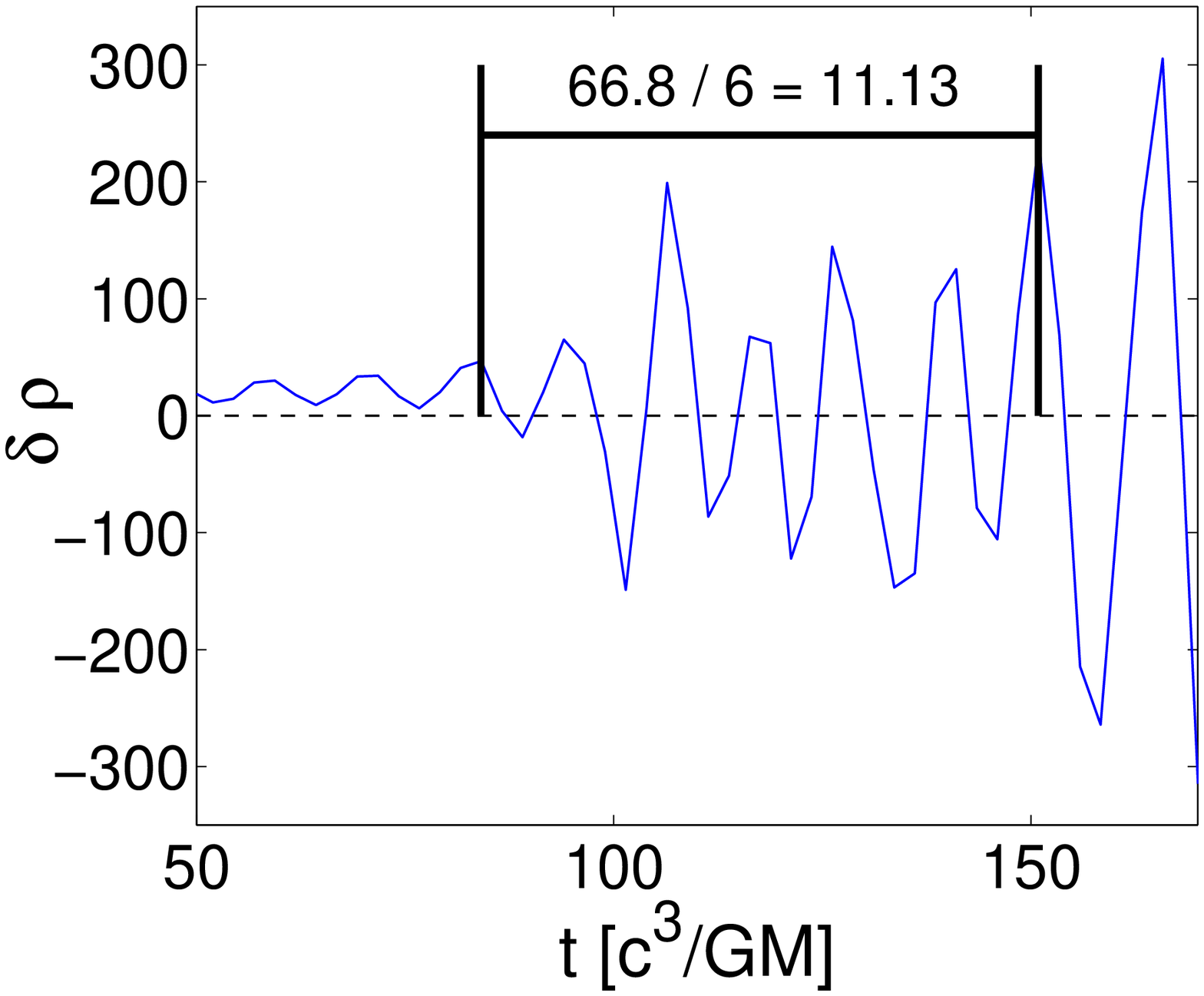}
\includegraphics[trim = 5mm 0mm 6mm 5mm, width = 0.242\textwidth]{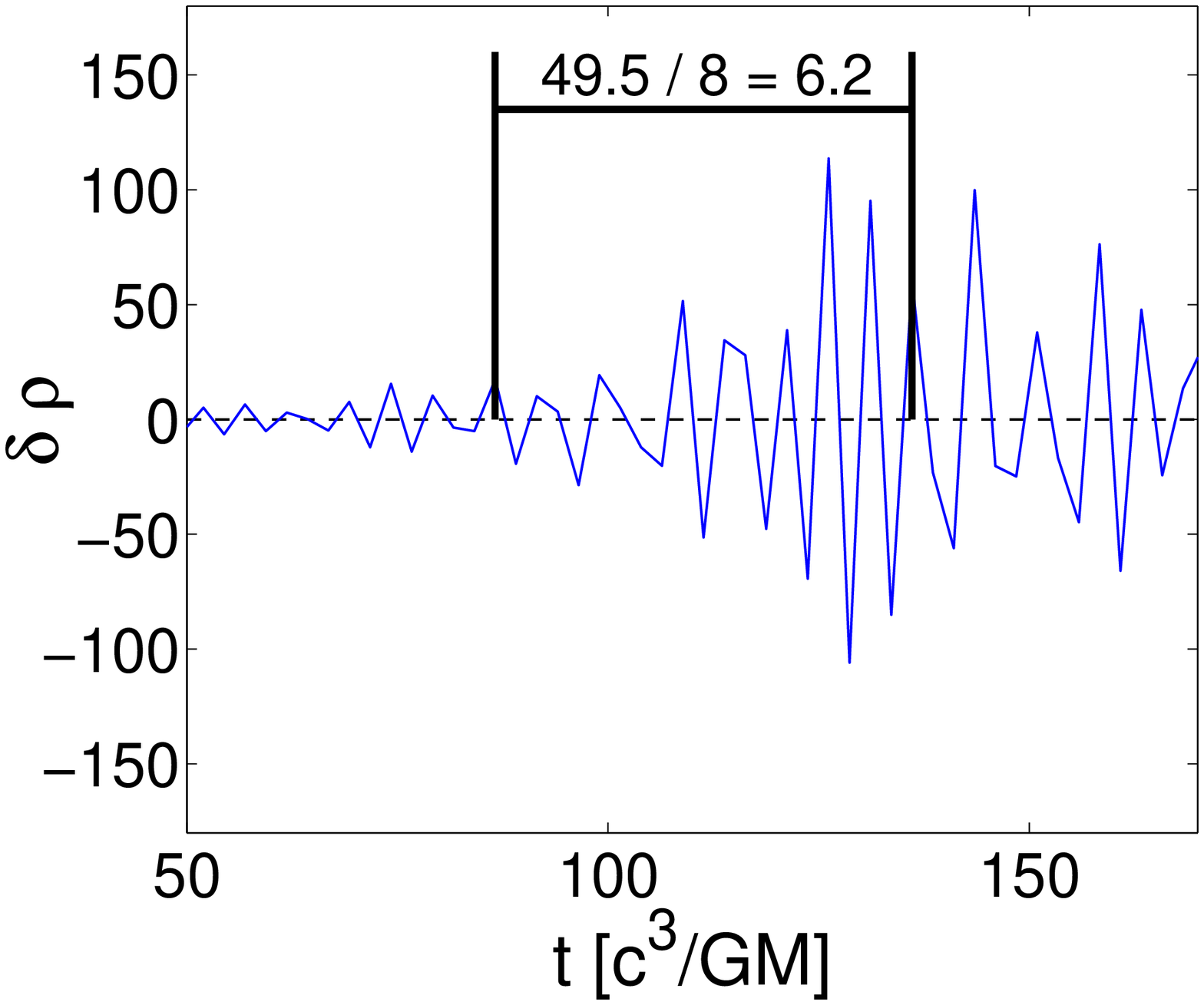}
\caption{Estimation of the perturbative oscillation period using the density data from the case A simulation. The $k_\phi = 4$ mode is presented on the {\em left}, with the $k_\phi = 8$ mode on the {\em right}.}
\label{fig:frequencies}
\end{figure}

Another way to check the consistency between our nonlinear GRMHD simulations and the linear perturbative analysis is to consider the real frequency (or period) of oscillations. In Fig. \ref{fig:frequencies}, we consider the real part of the density perturbation, $\delta \rho$, at $r = 3.5M$ for case A. Hence, this is the real part of the complex signal for which the modulus is shown in the last panel of Fig. \ref{fig:modesA}. From Fig. \ref{fig:frequencies}, we estimate the oscillation periods to be $T_4 = 11.1$ and $T_8 = 6.2$ (geometric units). The corresponding predictions from the linear analysis are $11.1$ and $5.53$, respectively, close enough to provide additional confidence that our GRMHD simulations are capturing the same modes as predicted by our perturbation analysis.

\section{Discussion and Conclusions}
\label{s.discuss}

In this paper, we have shown that strongly magnetized tori, such as the Komissarov solutions constructed in Section 2, are susceptible to a single unstable mode, corresponding to the non-axisymmetric MRI. Especially interesting is that this mode can not be stabilized by increasing the magnetic field strength. This is, at least partly, because, although $\beta$ can be made arbitrarily small by making $P_g$ small, $P_m$ can not be made arbitrarily large. Fundamentally, this is owing to the fact that the total pressure is required to go to zero at the torus surface, thus limiting the maximum pressure at the center. However, the torus does stabilize in the cold MHD limit, $P_g \rightarrow 0$ (or $c_s \rightarrow 0$), when the instability transforms into a co-rotation mode. We also demonstrated that the growth rate of this mode, as predicted by our linear perturbation analysis, matches well with that measured from our GRMHD simulations, at least in the cases where the growth is fast compared to the rate of background shearing. The simulations also confirmed that the growth rate scales with azimuthal wavenumber, as predicted by the linear perturbative analysis.

However, there are several curious features associated with this unstable mode, e.g., the zero amplitudes of their radial perturbations and the related independence of the growth rates on any components of the wave vector other than $k_\phi$. This is in disagreement with the general picture of the non-axisymmetric MRI given by \citet{Balbus92}. The discrepancy is possibly related to our neglect of the background shear, thereby enforcing a constant $k_r$.  However, only under such  an assumption were we able to deliver a simplified formula for the constant (local-in-time) growth rate. An accurate prediction of the growth rate using the \citet{Balbus92} approach would be much more difficult, since it would require an estimation of the full wave vector (not to mention the Newtonian character of their analysis). Regardless, we can say that the growth rates we find are significantly larger than those predicted in \citet{Balbus92}.  Whereas they mention growth rates of a few percent of the orbital frequency, we observe growth rates of order the orbital frequency or larger (e.g. for case B, with $k_\phi = 8$ at $r=3.5M$, the growth rate is $\approx \Omega$).  This is obviously aided by the linear dependence of our mode growth rates on $k_\phi$, a~feature that does not appear in the analysis of \citet{Balbus92}. The larger growth rates may be attributable to the highly relativistic background of the considered tori and the consequently smaller orbital frequencies.

While these issues may have some quantitative impact on our linear eigenanalysis, the results show remarkable agreement with the fully-nonlinear GRMHD simulations. Hence, we are confident in concluding that the Komissarov torus is prone to a local, non-axisymmetric magnetorotational instability that rapidly (i.e., on roughly the local orbital timescale) triggers the inward transport of mass (i.e., accretion).

In axisymmetry, on the other hand, we confirmed that the Komissarov torus is stable, even though this appears to be in conflict with \citet{Knobloch92}.  The stability criterion there (equation (9)) would appear to be violated for the Komissarov torus, yet we see no instability to axisymmetric perturbations in our linear eigenanalysis, nor in our 2D, axisymmetric GRMHD simulations.  Our speculation is that this discrepancy is attributable to the fact that the analysis of \citet{Knobloch92} ignores density gradients, which are important in the construction of finite tori and may contribute to their stability.

Finally, for completeness, we mention that the Komissarov torus is {\em not} unstable to the Parker interchange mode \citep{Parker66}.  The general criterion for that mode (at least for a plane parallel, isentropic atmosphere) is that instability results if the quantity $(b^\phi/\rho)^2$ (or $v_A/c_s$, where $v_A$ is the Alfv\'en speed) decreases vertically.  For the Komissarov torus, though, it is just the opposite, with $(b^\phi/\rho)^2$ being smallest in the equatorial plane and {\em increasing} vertically.  This explains why there is only one unstable mode (the non-axisymmetric MRI) in our analysis.

\section{Acknowledgements}

We thank M. Abramowicz, O. Blaes, J. Hor\'ak and W. Klu\'{z}niak for extensive discussions and comments. Research supported in part by the Polish NCN grant UMO-2011/01/B/ST9/05439 as well as the National Science Foundation grants NSF AST-1211230 and NSF PHY11-25915.

\bibliographystyle{mn2e}
\bibliography{refs}

\appendix

\section{Perturbed system of ideal MHD equations}
\label{ap.system}

Below we give a system of perturbed MHD equations for the \citet{Komissarov06} solution. Note that the equations are \textit{not} valid for the general flow, in particular terms such as $\partial_\alpha u^\alpha$ vanish because of the assumed symmetries. To simplify the equations, the following substitutions are used
\be
h = w + b^2 ~,
\ee 
\be 
\delta P_g = K_g w^{\Gamma-1} \delta w ~,
\ee
\be
\delta \rho = \left(1 - K_g \frac{\Gamma^2}{\Gamma-1}w^{\Gamma-1} \right) \delta w ~.
\ee
We also put $k_t \equiv \omega$ in order to fully benefit from the Einstein summation convention. Hence, the $\alpha$ and $\beta$ indices denote summation over all four spacetime dimensions.

The perturbed continuity equation reads
\be 
0 = \partial_r(\sqrt{-g} \rho)  \delta u^r  -i\sqrt{-g} k_\alpha \left( u^\alpha  \delta \rho + \rho  \delta u^\alpha \right) ~.
\ee
From the induction equations (radial, vertical, and azimuthal) we find
\be 
0 = k_\alpha \left( u^\alpha \delta b^r  - b^\alpha \delta u^r \right) ~,
\ee
\be 
0 = k_\alpha \left( u^\alpha \delta b^\theta  - b^\alpha \delta u^\theta \right) ~,
\ee
\begin{eqnarray}
0 &=& \partial_r(\sqrt{-g} b^\phi ) \delta u^r -\partial_r (\sqrt{-g} u^\phi) \delta b^r \nonumber \\ 
&-& i\sqrt{-g} k_\alpha \left( u^\alpha \delta b^\phi + b^\phi \delta u^\alpha - u^\phi \delta b^\alpha - b^\alpha \delta u^\phi \right) ~.
\end{eqnarray}
Finally, the perturbed Euler equations (vertical, azimuthal, and time) read
\be 
0 = g_{\theta \theta} k_\alpha \left( h u^\alpha \delta u^\theta - b^\alpha \delta b ^\theta \right) + k_\theta (\delta P_g + b_\alpha \delta b^\alpha ) ~,
\ee
\begin{eqnarray}
0 &=& \partial_r (\sqrt{-g} h u_t) \delta u^r - \partial_r(\sqrt{-g} b_t) \delta b^r \nonumber \\
&-& i \sqrt{-g} k_\alpha \left[ h \left( u_t \delta u ^\alpha + u^\alpha g_{t \beta} \delta u^\beta \right) - \left( b_t \delta b^\alpha + b^\alpha g_{t \beta} \delta b^\beta \right) \right] \nonumber \\
&-& i \sqrt{-g} \left[ k_t \left( \delta P_g + b_\alpha \delta b^\alpha \right) + u_t k_\alpha u^\alpha \left( \delta w + 2 b_\beta \delta b^\beta \right) \right] ~,
\end{eqnarray}
\begin{eqnarray}
0 &=& \partial_r (\sqrt{-g} h u_\phi) \delta u^r - \partial_r(\sqrt{-g} b_\phi) \delta b^r \nonumber \\
&-& i \sqrt{-g} k_\alpha \left[ h \left( u_\phi \delta u ^\alpha + u^\alpha g_{\phi \beta} \delta u^\beta \right) - \left( b_\phi \delta b^\alpha + b^\alpha g_{\phi \beta} \delta b^\beta \right) \right] \nonumber \\
&-& i \sqrt{-g} \left[ k_\phi \left( \delta P_g + b_\alpha \delta b^\alpha \right) + u_\phi k_\alpha u^\alpha \left( \delta w + 2 b_\beta \delta b^\beta \right) \right] ~.
\end{eqnarray}
Note that we did not use the radial Euler equation, which is particularly complicated because of the lack of a Killing symmetry. Thus, we close the system with the perturbed 4-velocity normalization:
\be
0 = u_t  \delta u^t + u_\phi \delta u^\phi ~.
\ee
Additionally, we have a condition derived from the orthogonality of $u^\alpha$ and $b^\alpha$:
\be
0 = u_t \delta b^t + u_\phi \delta b^\phi + b_t \delta u^t + b_\phi \delta u^\phi ~.
\ee
The system used for the numerical eigenanalysis is given by equations (A4)-(A12), though (A11) is only used to eliminate $\delta u^t$.
To calculate the necessary radial derivatives, we use the numerically evaluated values of $\partial_r w$ and $\partial_r \ell$. The system of equations (A4)-(A12) can now be evaluated at a given radius for the chosen steady-state solution parameters and cast into the form discussed in Section \ref{s.perturb}, which can then be solved for the generalized eigenvalues.

\end{document}